\documentclass[a4paper,fleqn,usenatbib]{mnras}

\usepackage{amssymb}
\usepackage{graphicx}
\usepackage{longtable}
\usepackage[T1]{fontenc}
\usepackage{ae,aecompl}

\def\Teff{$T_{\rm eff}$}
\def\logg{$\log\,g$}

\def\Vt{V$_{\rm t}$}

\def\aj{AJ}%
\def\araa{ARA\&A}%
\def\apj{ApJ}%
\def\apjl{ApJ}%
\def\apjs{ApJS}%
\def\ao{Appl.~Opt.}%
\def\aap{A\&A}%
\def\aaps{A\&AS}%
\def\jcap{J.~Cosm.~Astr.~ Phys.}%
\def\mnras{MNRAS}%
\def\pasa{PASA}%
\def\prd{Phys.~Rev.~D}%
\def\pasj{PASJ}%
\def\solphys{Sol.~Phys.}%
\def\physrep{Phys.~Rep.}%
\newcommand {\apgt} {\ {\raise-.5ex\hbox{$\buildrel>\over\sim$}}\ }
\newcommand {\aplt} {\ {\raise-.5ex\hbox{$\buildrel<\over\sim$}}\ }

\title[The metal-poor solar neighbourhood]
{Observing the metal-poor solar neighbourhood: a comparison of galactic chemical evolution predictions
\thanks{Based on observations collected at OHP observatory, France}
\thanks{Table \ref{EWline} are only available in electronic form}
}
\author[T.~Mishenina  et al.]
{T.~Mishenina$^{1}$,
M.~Pignatari$^{2,3,6}$,
B.~C\^ot\'e$^{3,4,5,6}$
F.-K.~Thielemann$^{7}$,
 C.~Soubiran$^{8}$,\newauthor
 N.~Basak$^{1}$,
 T.~Gorbaneva$^{1}$,
S.A.~Korotin$^{1,9}$,
V.V.~Kovtyukh$^{1}$,
B.~Wehmeyer$^{7}$,\newauthor
S.~Bisterzo$^{3,10,11}$,
C.~Travaglio$^{3,10,11}$,
B.K.~Gibson$^{2,6}$,
C.~Jordan$^{2,6}$,
A.~Paul$^{4}$,
C.~Ritter$^{3,6}$\newauthor
F.~Herwig$^{3,6}$
 \\
$^{1}$Astronomical Observatory, Odessa National University,         and \\
    Isaac Newton Institute of Chile, Odessa branch,
       Shevchenko Park, 650014, Odessa, Ukraine\\
$^{2}$ E.A. Milne Centre for Astrophysics, University of Hull, HU6~7RX, United Kingdom\\
$^{3}$ The NuGrid Collaboration, http://www.nugridstars.org\\ 
$^{4}$ Department of Physics and Astronomy, University of Victoria, Victoria, BC, V8W 2Y2, Canada\\
$^{5}$ National Superconducting Cyclotron Laboratory, Michigan State University, East Lansing, MI, 48824, USA \\ 
$^{6}$ Joint Institute for Nuclear Astrophysics Center for the Evolution of the Elements, USA \\
$^{7}$ Department of Physics, University of Basel, Klingelbergstrabe 82,
        4056 Basel, Switzerland\\
$^{8}$  Laboratoire d'Astrophysique de Bordeaux, 
        Univ. Bordeaux  - CNRS, B18N,  all\'ee Geoffroy Saint-Hilaire, 33615 Pessac, France\\
$^{9}$ Crimean Astrophysical Observatory, Nauchny, 298409, Crimea\\
$^{10}$ INAF, Astrophysical Observatory Turin, Strada Osservatorio 20, I-10025 Pino Torinese (Turin), Italy \\
$^{11}$ B2FH Association, Turin, Italy \\}

\date{Accepted 2017 xxx. Received 2017 xxx; in original form 2017 xxx}
\pubyear{2017}

\begin{document}

\label{firstpage}

\pagerange{\pageref{firstpage}--\pageref{lastpage}}

\maketitle

\begin{abstract}
Atmospheric parameters and chemical compositions for ten stars with 
metallicities in the region of $-2.2<$[Fe/H]$<-0.6$ were 
precisely determined using high resolution, high signal to noise, spectra.
For each star the abundances, for 14 to 27
elements, were derived using both LTE and NLTE approaches. In particular, differences by assuming LTE or NLTE are about 0.10 dex; depending on [Fe/H], \Teff, gravity and element lines used in the analysis.
We find that the O abundance has the largest error, ranging from 0.10 and 0.2 dex. The best measured elements are Cr, Fe, and Mn; with errors between 0.03 and 0.11 
dex. The stars in our sample were included in previous different observational work. 
We provide a consistent data analysis. The data dispersion introduced in the literature 
by different techniques and assumptions used by the different authors is within the observational errors, excepting
for HD103095. 
We compare these results with stellar observations from 
different data sets and a number of theoretical galactic chemical 
evolution (GCE) simulations. We find a large scatter in the GCE 
results, used to study the origin of the elements. Within this scatter
as found in previous GCE simulations, we cannot reproduce the evolution of the 
elemental ratios [Sc/Fe], [Ti/Fe], and [V/Fe] at different metallicities. 
The stellar yields from core collapse supernovae (CCSN) are likely primarily responsible 
for this discrepancy. Possible solutions and open problems are discussed.
\end{abstract}


\section{Introduction}

The observation of chemical abundances in stars at different metallicities provides a fundamental tool to study the evolution of our Galaxy \citep[e.g.,][]{reddy:03, reddy:06, frebel:10, yong:13, battistini:16}. 
Determination of parameters and chemical compositions of stars with low metal abundances 
is more challenging compared to stars of solar metallicity. 
This is due to the influence of metallicity on atmospheric parameters,  
caused primarily by  
stronger deviations from Local Thermodynamic Equilibrium (LTE).
This is associated with a decrease in electron density and reduction of 
collisions in reaching equilibrium \citep[e.g.,][]{mash:00}.
In this work we analyse ten stars   
which have been investigated by previous studies, and estimate the accuracy of determination of parameters and chemical compositions.
The stars cover a metallicity range $-2.2<$ [Fe/H] $<-0.6$. Stars within this range reveal crucial insights about the chemical evolution of the Galaxy. It includes the region [Fe/H] $\lesssim -1$, where typically only massive stars and super Asymptotic Giant Branch (AGB) stars have sufficient time to contribute significantly to the chemical enrichment history of the Galaxy \citep[e.g.,][]{nomoto:13}. Where [Fe/H] $\gtrsim -1$ the contribution from lower mass AGB stars and supernovae type Ia (SNe Ia) affects the chemical enrichment history in the galactic disk \citep[e.g.,][]{matteucci:85}.
With this work, we aim to provide new observational data to study this metallicity region, with special attention to observational uncertainties.
Most of the stars analysed have been included in previous works from other authors, and included in large stellar compilations. Relevant differences exist between different measurements for some cases. Such differences are due to legitimate assumptions and choices made by the authors. 
Our results are compared with a number of galactic chemical evolution (GCE) simulations. Adopting the same approach used for the observational analysis, the different simulations are discussed, where the results are a product of the theoretical setups adopted by the authors.

The paper is organised as follows: 
the observations and selection of stars, and definition of the main stellar 
parameters are described in \S \ref{sec: stellar param}; ages and kinematic parameters are presented in \S \ref{sec: ages, kinematics}; the selection of lines is given in \S \ref{sec: select lines}; the abundance determinations and the error analysis are presented in 
\S \ref{sec: abundance determination}. 
Results, membership of galactic structures and comparison with other data and with theoretical GCE simulations are given in \S \ref{sec: result, discuss}. 
Conclusions are drawn in \S \ref{sec: conclusions}.

\section{Observations, selection and parameters of the stars}
\label{sec: stellar param}

For this study ten metal-poor stars with different metallicities from $-0.6$ to $-2.2$ were selected. Their spectra were obtained with the SOPHIE echelle spectrograph \citep{perruchot:08} attached to the 1.93 m telescope at the Observatoire de Haute Provence, France. The resolving power of the spectrograph is R = 75 000, the spectra are in the  wavelength range  $\lambda$~4400--6800\,\AA\, and signal-to-noise ratio of about 100--400. The list of target stars, observation dates,  signal-to-noise ratios and the radial velocity RV are given in Table \ref{observ}.

\begin{table}
\caption{Observation data for our target stars.}
\label{observ}
\begin{tabular}{lccr}
\hline
\hline
HD     &  	date  &  S/N & RV, km/s \\
\hline
  6582&	 	05.12.2013&		418 & --96.305   \\
  6833&		19.09.2006&		234 & --243.410   \\
 19445&		 17.01.2010&		102 & --139.936    \\
 22879&		06.12.2013&		207 & 120.397   \\
 84937&	 	09.12.2013&		167 & --15.015  \\
103095&	  	06.12.2013&		259 & --97.922    \\
170153&		30.08.2011&		317 &  37.781  \\
216143&	 	19.09.2006&		153 & --116.462  \\
221170&		19.09.2006&		201 &  --121.717   \\
224930&     16.01.2011& 326& --41.105    \\
\hline                                                                                             
\end{tabular}                                                                                        
\end{table}

\noindent
The observations were retrieved from the on-line SOPHIE archive\footnote{http://atlas.obs-hp.fr/sophie/} which provides science-ready spectra with cross-correlation functions and radial velocity measurements.
Further spectra  processing (the continuum placement,  equivalent width (EW) measurements, etc.) was conducted using the DECH20 software package by \citet{galazut:92}. 

\subsection{Effective temperature \Teff}
	
The main methods to calculate \Teff\ 
are based on photometric calibrations and on spectroscopic calibrations using Fe abundance lines; assuming the absence of any relationship between the elemental abundance  
estimated by a certain spectral line and the lower excitation potential $E_{low}$ of the line for a given temperature. In this study we applied the colour-\Teff\ calibrations of the B--V and b--y colour indices for dwarfs \citep[][]{alonso:96} and giants \citep[][]{alonso:99}, 
taking into account the stellar metallicity.   

The B--V and b--y data were taken from the SIMBAD database.
The \Teff\ determinations for different values of the B--V and b--y colour indices for our stars, and the \Teff\ values obtained using spectroscopic methods are presented in Table \ref{param}. 
Figure \ref{elowew} shows the dependence of the iron abundance log A(Fe~{\sc i})  on the lower excitation potential $E_{low}$ for each target star,  
where an abundance of the hydrogen is log A(H) = 12.0.
 
By using different colour indices (B--V or b--y), there is an average variation of \Teff\ of 50 K, with the maximum difference never exceeding 100 K (see Table \ref{param}).
The comparison of these results with the spectroscopic \Teff\ 
gives higher discrepancies ($\Delta$\Teff\ $>$ 100 K), 
particularly when using the B--V colour index. 
We find the opposite situation for star HD6833, as the b--y colour index of \Teff\ determination results in higher difference between the \Teff\ values. However for the star HD224930 both colour indices give $\Delta$\Teff\ $>$ 100 K.

\begin{table*}
\caption{Parameters of studied stars.}
\label{param}
\begin{tabular}{lccccccccccccc}
\hline
\hline
HIP   &	HD     & V     &B-V    & b-y   &c1    &	[Fe/H]& \Teff\ sp &\Teff\ (B-V) &\Teff\ (b-y) & log$_{P}$ & log$_{IE}$ & \Vt & \Vt (S)\\ 
      &	       &       &       &       &      &	      & K         &K            &K            &           &            & km/s& km/s   \\ 
\hline
5336  &	6582&	5.17&	0.70&	0.437&	0.213&	--0.83&	5350&	5241&	5336	&	4.56&	4.50 &0.4&0.80	 \\ 
     &	&	&	&	0.441&	0.208&	&	--&	&	5311	&	&	&	 \\ 
5458  &	6833   & 6.74&	1.17&	0.753&	0.487&	--0.77&	4415&	4382&	4309	&	1.79&	1.50 & 1.3&1.44	 \\ 
14594 &	19445&	8.06&	0.45&	0.352&	0.208&	--2.16&	5830&	5923&	5892	&	4.45&	4.00&1.1&1.20	 \\ 
17147 &	22879&	6.69&	0.54&	0.365&	0.272&	--0.91&	5825&	5793&	5804	&	4.29&	4.42&0.9&1.09	 \\ 
      &	&	&	&	0.369&	0.302&	&	--&	&	5867	&	&	 &	 \\ 
48152 &84937&	8.32&	0.41&	0.293&	0.390&	--2.24&	6325&	6084&	6429	&	4.18&	3.95&1.4&1.49	 \\ 
      &	&	&	0.37&	0.302&	0.369&	&	--&	6259&	6349	&	&	 &	 \\ 
57939 &103095&	6.45&	0.74&	0.484&	0.155&	--1.35&	5100&	5023&	5035	&	4.88&	4.65&0.4&0.56	 \\ 
      &	&	&	0.76&	0.487&	0.151&	&	--&	4966&	5017	&	&	&	 \\ 
      &	&	&	&	0.475&	0.196&	&	--&	&	5103	&	&	 &	 \\ 
89937 &	170153&	3.58&	0.50&	&	&	--0.61&	6170&	6000&		&	4.17&	4.25&0.7&1.38	 \\ 
      &	&	&	0.48&	&	&	&	--&	6083&		&	&	 &	 \\ 
112796&	216143&	7.81&	&	0.690&	0.572&	--2.26&	4455&	&	4471	&	0.92&	1.05& 2.0&1.69	 \\ 
      &	&	&	&	0.681&	0.558&	&	--&	&	4491	&	&	 &	 \\ 
      &	&	&	0.97&	&	&	&	--&	4357&		&	&	 &	 \\ 
115949&	221170&	7.66&	1.08&	0.747&	0.564&	--2.26&	4415&	4403&	4354	&	1.89&	1.05&1.9&1.67	 \\ 
      &	&	&	0.99&	0.756&	0.556&	&	--&	4354&	4337	&	&		& \\ 
171   &	224930&	5.75&	0.66&	0.432&	0.218&	--0.79&	5500&	5382&	5371	&	4.44&	4.40&0.3&0.93	 \\ 
      &	&	&	0.67&	0.435&	0.225&	&	--&	5350&	5358	&	&		 \\ 
\hline
\end{tabular}  
Notes: The values of \Vt (S) are calculated by using the parametric formula by \citep{sitn:15}.
\end{table*}

It is important to account for the reddening of E(B--V) in \Teff , and other parameter determinations based on  photometric calibrations. Our investigated stars are in close vicinity to the Sun meaning most of them have little reddening. Accounting for the reddening for more distant stars, like the star HD221170 in our sample, can significantly change \Teff. For instance in \cite{ivans:06} for the star HD221170 it was shown that usage of larger  E(B--V) value of reddening resulted in increased temperature (4610 K). At this temperature the authors observe the dependence of the iron abundance on the lower excitation potential $E_{low}$ for a given line, which should not be if the \Teff\ is correctly defined. 
Given these uncertainties we opted to use the spectroscopic method for the temperature determination (see Figure \ref{elowew}). We are aware that the accuracy of this method depends on the oscillator strengths used in the calculations with allowance of deviations from LTE (NLTE). As reported in the studies by  \cite{mash:11}  and  \cite{sitn:15} for the stars with [Fe/H] $>$ $-$1 and effective temperatures up to 5800 K, the NLTE corrections for neutral iron are smaller than 0.05 dex and they increase with decreasing metallicity. This study also shows deviations from LTE for iron do not exceed 0.05 dex for our target stars with [Fe/H] of about $-$2.25 dex and \Teff\  $<$ 5000 K. Only for the stars HD84937 and HD19445 do these deviations reach  of order 0.05--0.12 dex; depending on the excitation potential of the employed iron lines the deviations from LTE decrease with increasing excitation potential \citep{berg:12}. Most neutral iron lines used in our calculations have the lower excitation potential more than 2 eV. However, the estimated effect of deviations from LTE on the neutral iron lines varies significantly across different studies. This is due to the complexity of the multi-level model of an iron atom, which requires a large amount of atomic data (for which there are high uncertainties). That is one of the arguments in favour of application of the LTE analysis. Additionally, the majority of metal-poor stars' chemical composition estimates, and their in-depth study, have been performed under the assumption of LTE for parameter and metallicity determinations.

\subsection{Surface gravities \logg}
 
For the target stars we used two methods for determination of \logg: 
1) standard formula using the parallax (P): \\
\logg$_{P}$ = --12.50+log(M/M$_\odot$)+4\Teff\ +0.4(Mv+BC), \\
where 
 M/M$_\odot$  - mass of the star in units of solar masses,
$ Mv $ - absolute magnitude,
$ BC $ - bolometric correction;
bolometric corrections are taken from \cite{flower:75}, absolute
magnitude is determined by the parallax P from the catalog Hipparcos \citep{leeuw:07}; 
and 2) iron ionisation equilibrium (IE, spectroscopic method) of the neutral 
and ionised iron. This method implies that similar abundances are obtained from 
the neutral iron Fe~{\sc i} and ionised iron Fe~{\sc ii} lines.

Table \ref{param} presents the \logg\  determinations by the two methods, namely \logg$_{P}$ and \logg$_{IE}$,  respectively.  We obtained a good agreement between  the \logg\ definitions with these two methods, except for two stars, namely HD19445 and HD221170. In our opinion this is due to taking into account the reddening at the gravity definitions using parallax. For instance, accounting for the error in the parallax determination for HD221170 resulted in \logg\ from 1.90 to 1.48, while factoring in the E(B-V) reduced the value of \logg\ down to 1.66 with the parallax P = 0.00294. When determining \logg\ using the parallax \logg$_{P}$, the primary error is introduced by the accuracy of the parallax itself and by the accounting for the reddening; when using spectroscopic method for determination of \logg\ (\logg$_{IE}$), the use of oscillator strengths and accounting for deviations from LTE are essential. However, as shown in 
\citet{berg:12}, a small NLTE correction is needed to establish ionisation equilibrium at solar metallicities, while for very metal-poor stars these effects reach only of +0.1 dex on Fe~{\sc i} lines. Fe~{\sc ii} lines are basically not affected by departures from LTE. 
Since the effect of NLTE on \logg\ determination is rather small \citep[see also][]{jofre:14}, in this study we used the spectroscopic determinations of \logg.

\subsection{Turbulent velocity \Vt}

\begin{figure*}
\begin{tabular}{cc}
\includegraphics[width=8.4cm]{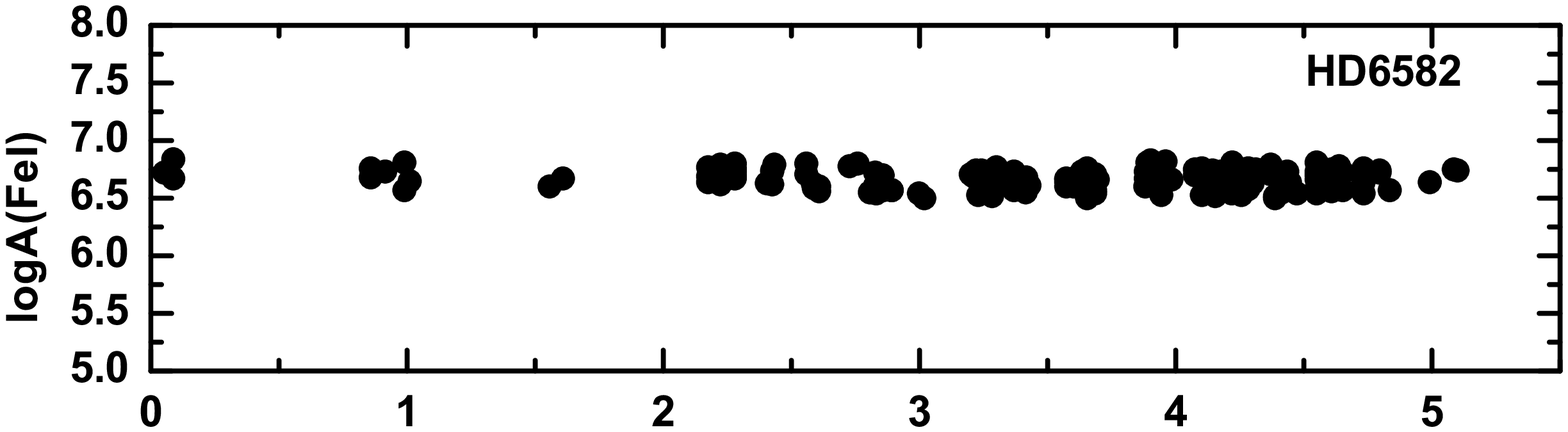}  &\includegraphics[width=8.4cm]{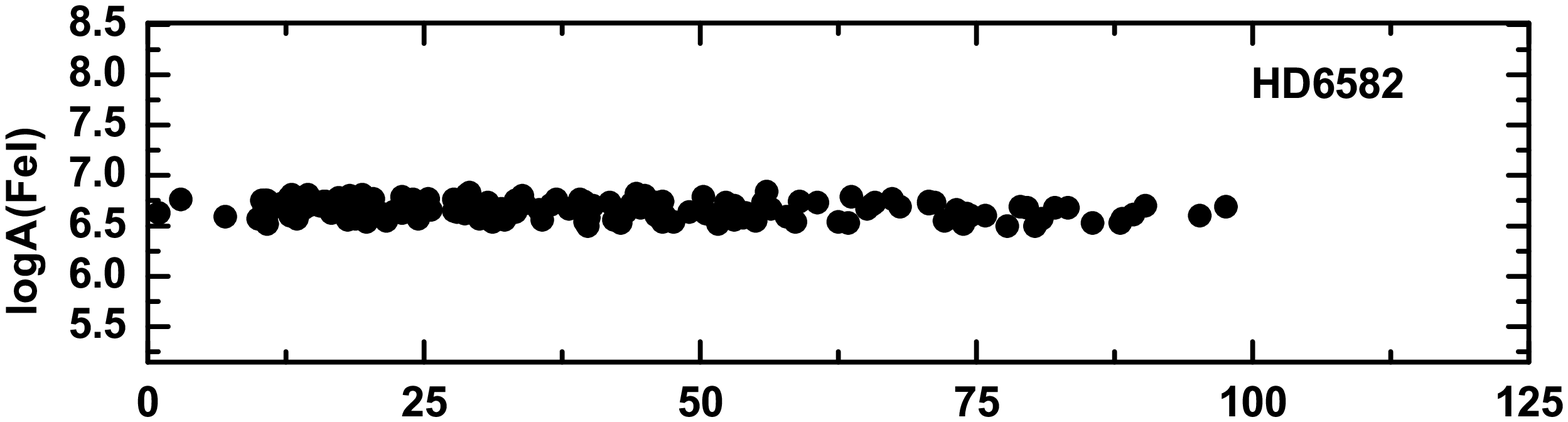}\\  
\includegraphics[width=8.4cm]{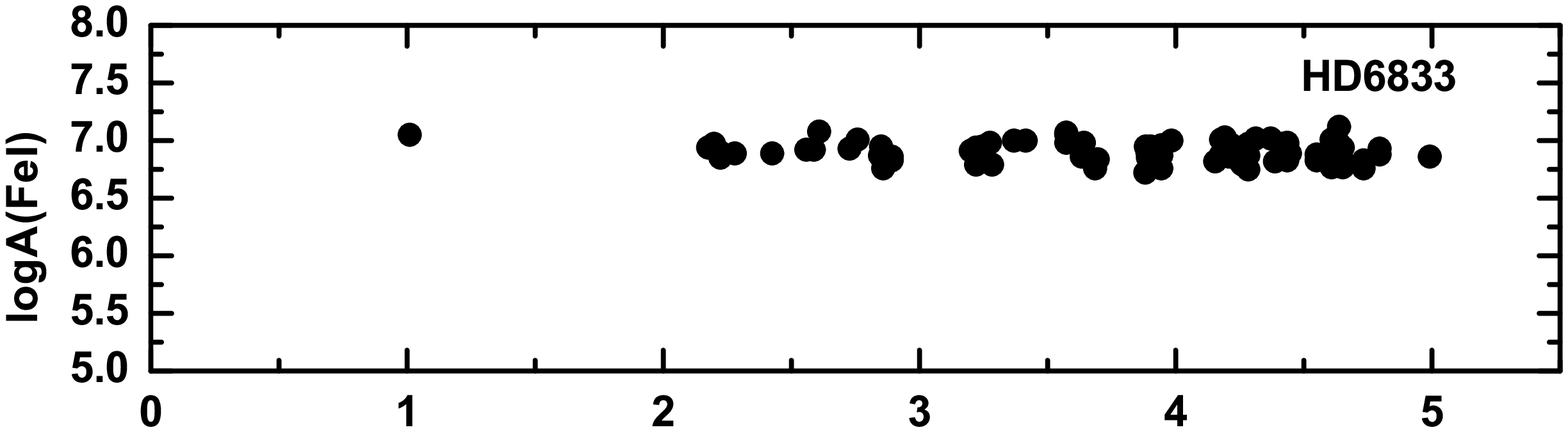}  &\includegraphics[width=8.4cm]{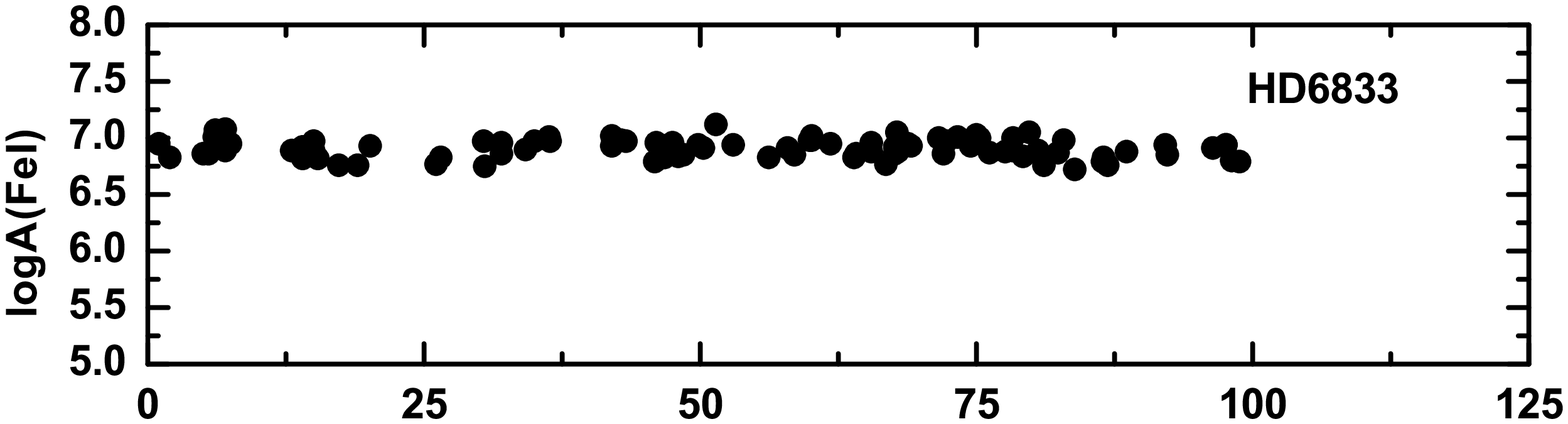}\\  
\includegraphics[width=8.4cm]{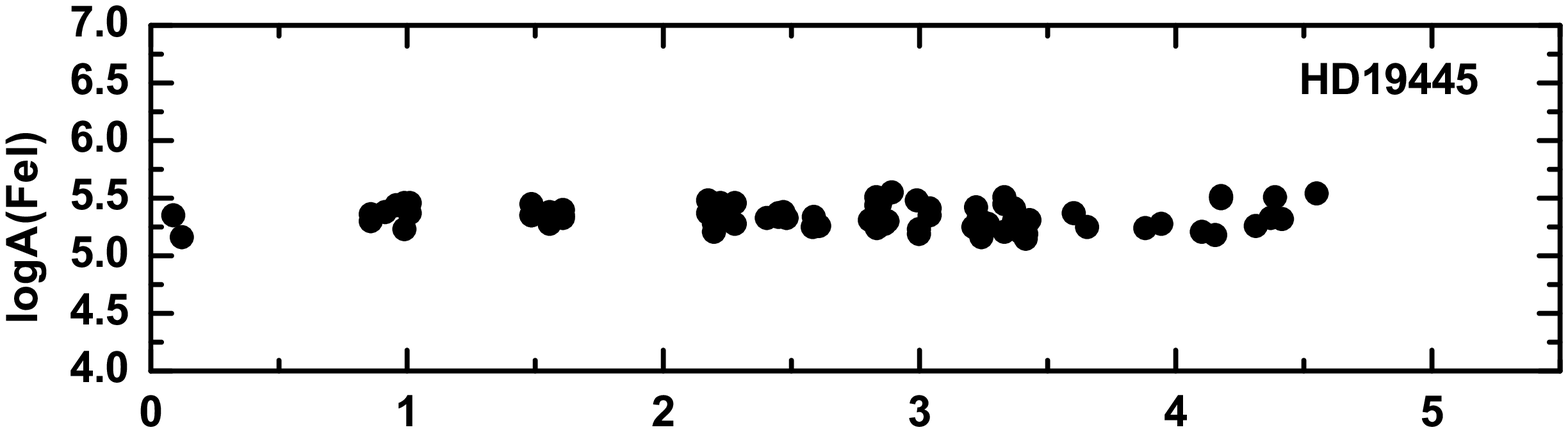} &\includegraphics[width=8.4cm]{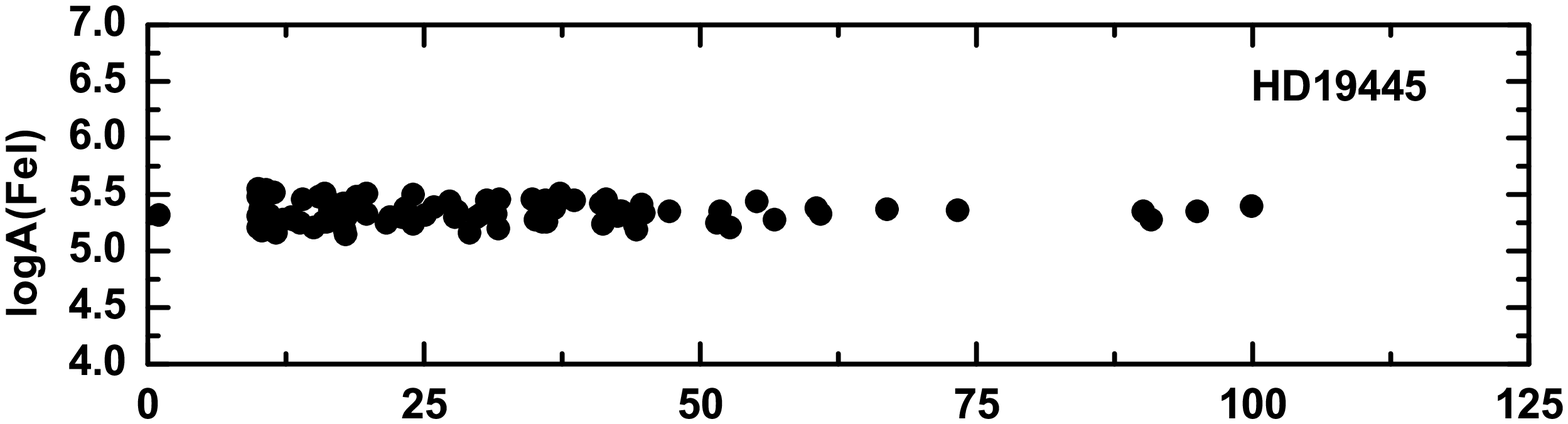}\\ 
\includegraphics[width=8.4cm]{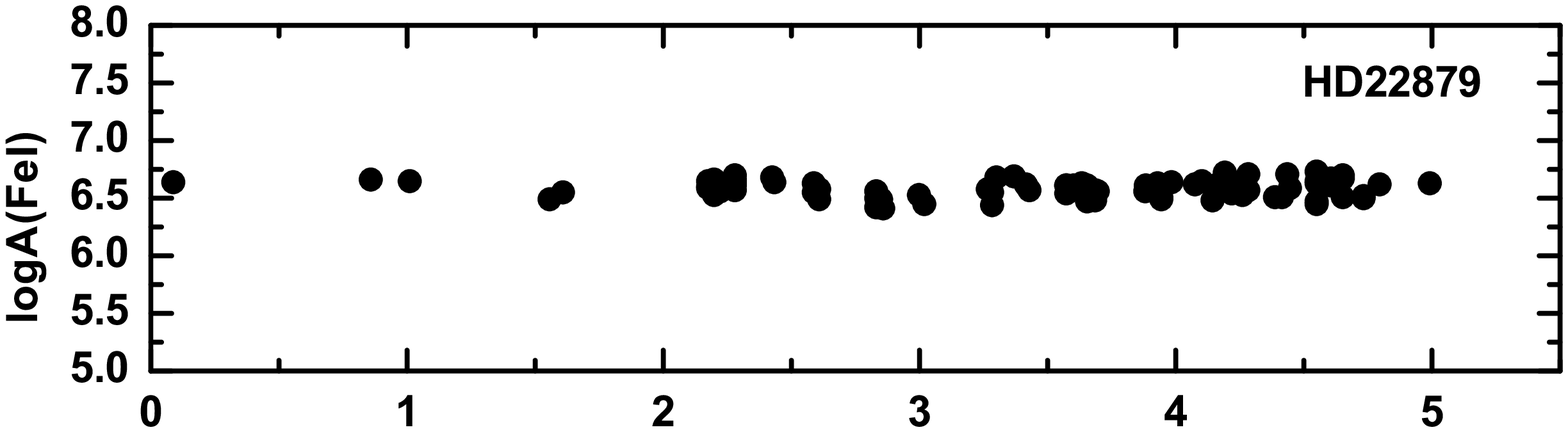} &\includegraphics[width=8.4cm]{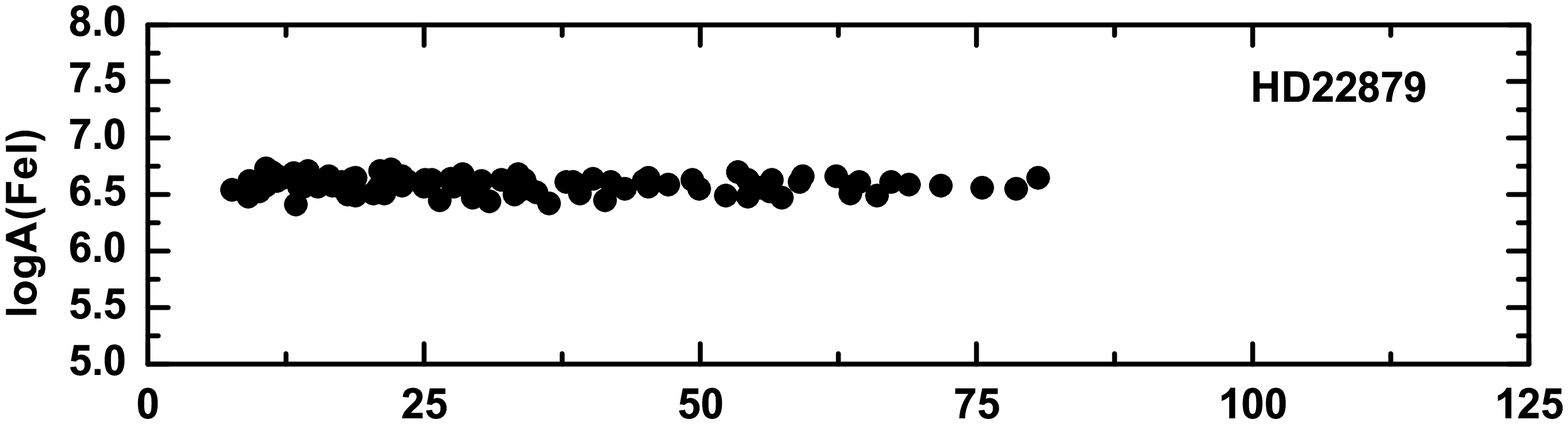}\\ 
\includegraphics[width=8.4cm]{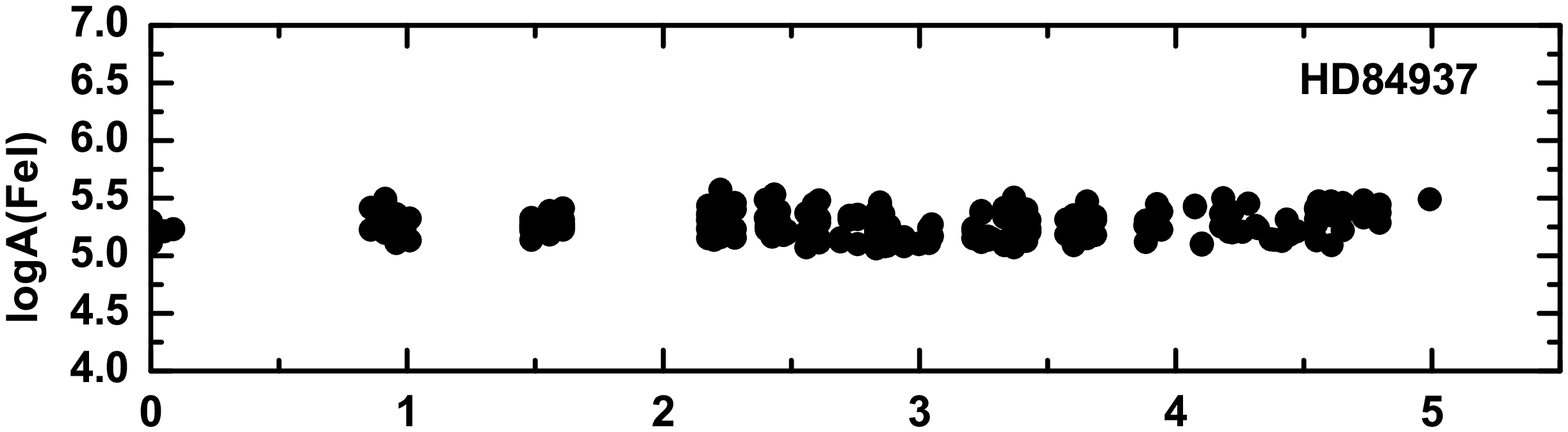} &\includegraphics[width=8.4cm]{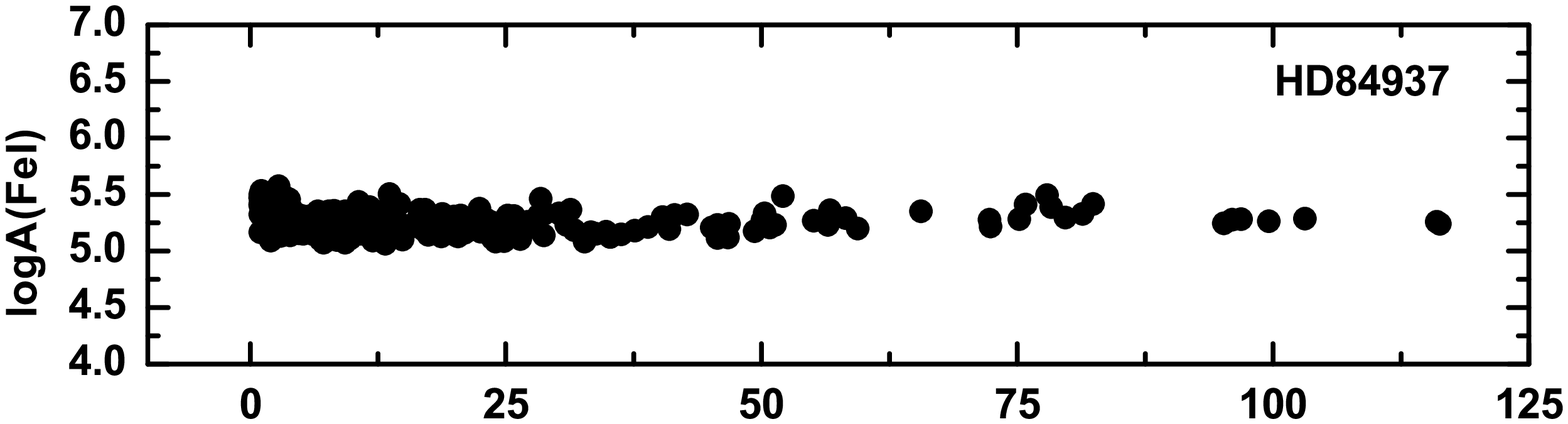}\\ 
\includegraphics[width=8.4cm]{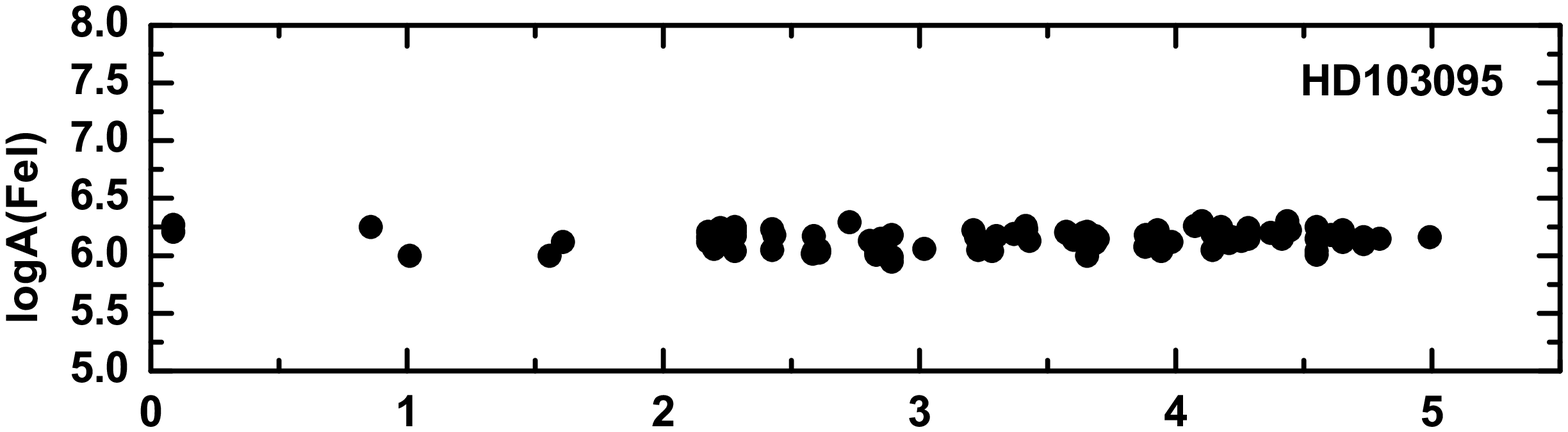}&\includegraphics[width=8.4cm]{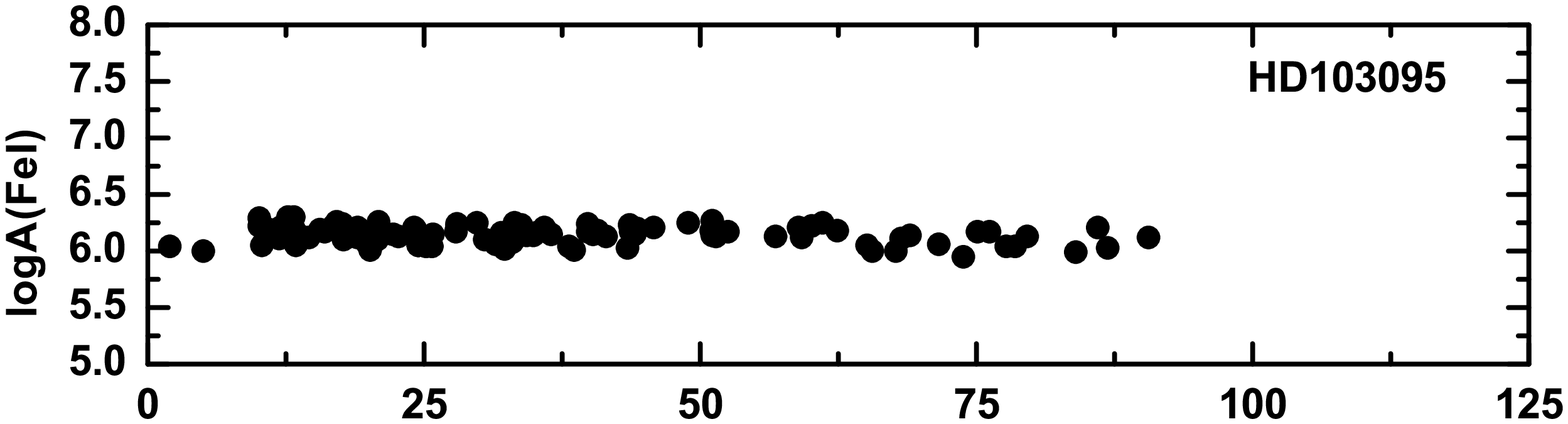}\\
\includegraphics[width=8.4cm]{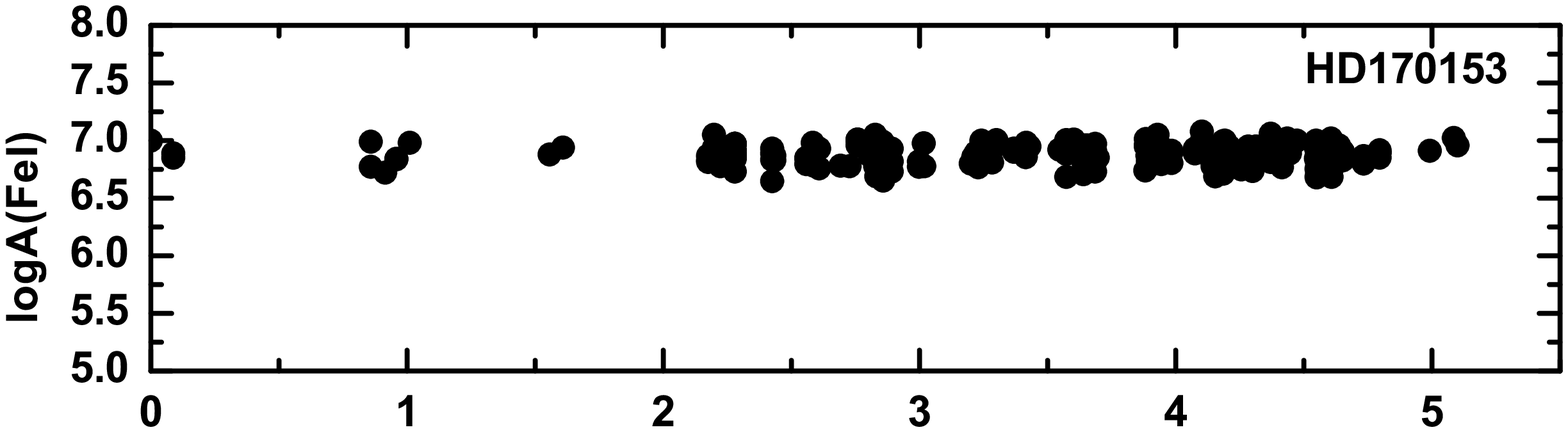}&\includegraphics[width=8.4cm]{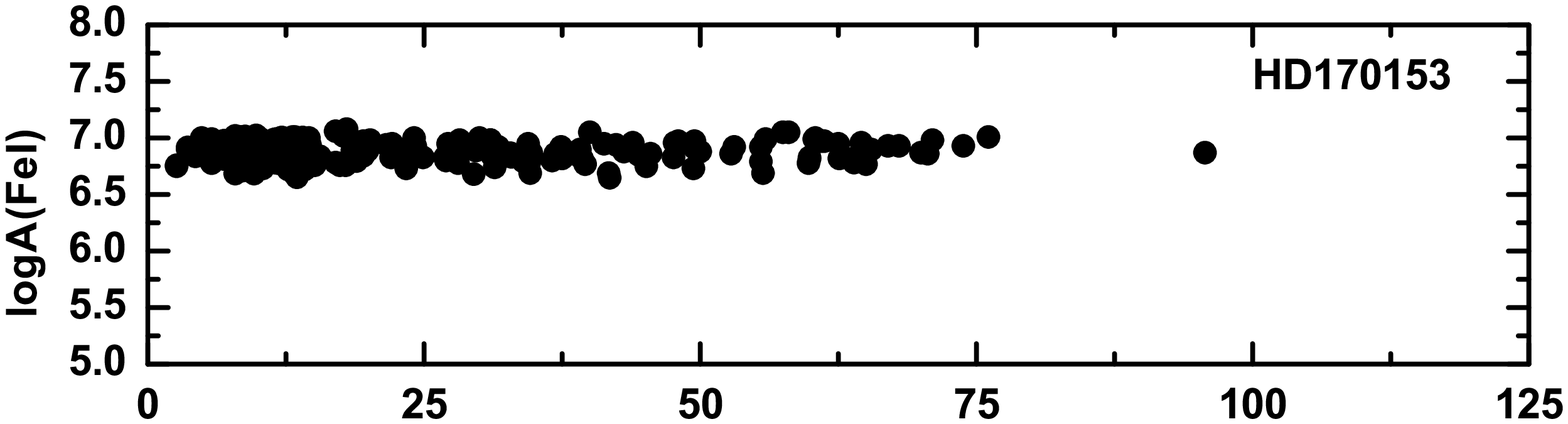}\\
\includegraphics[width=8.4cm]{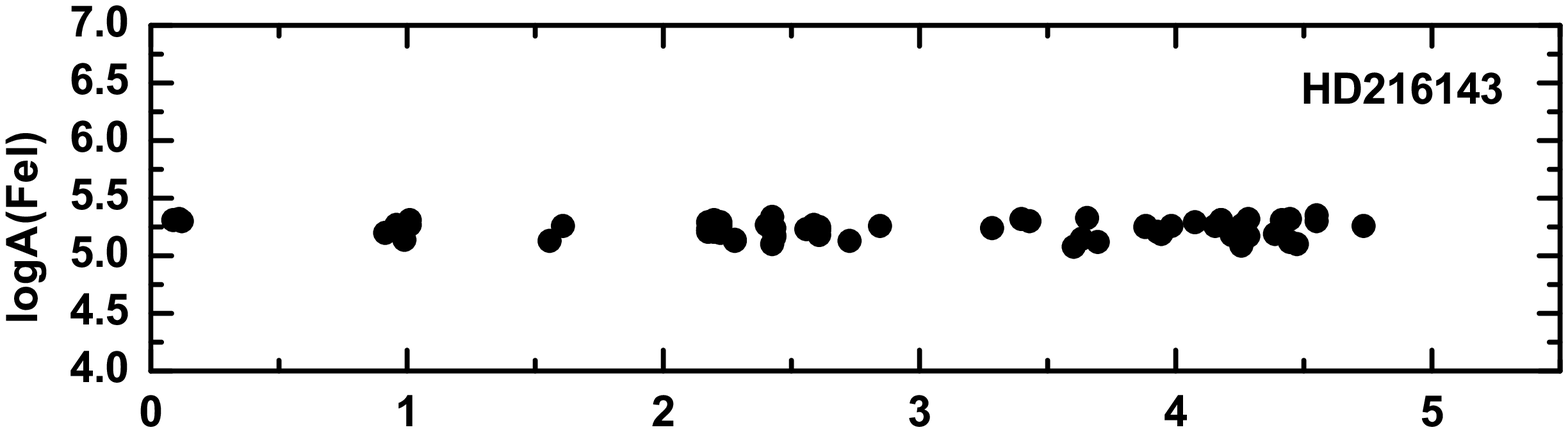}&\includegraphics[width=8.4cm]{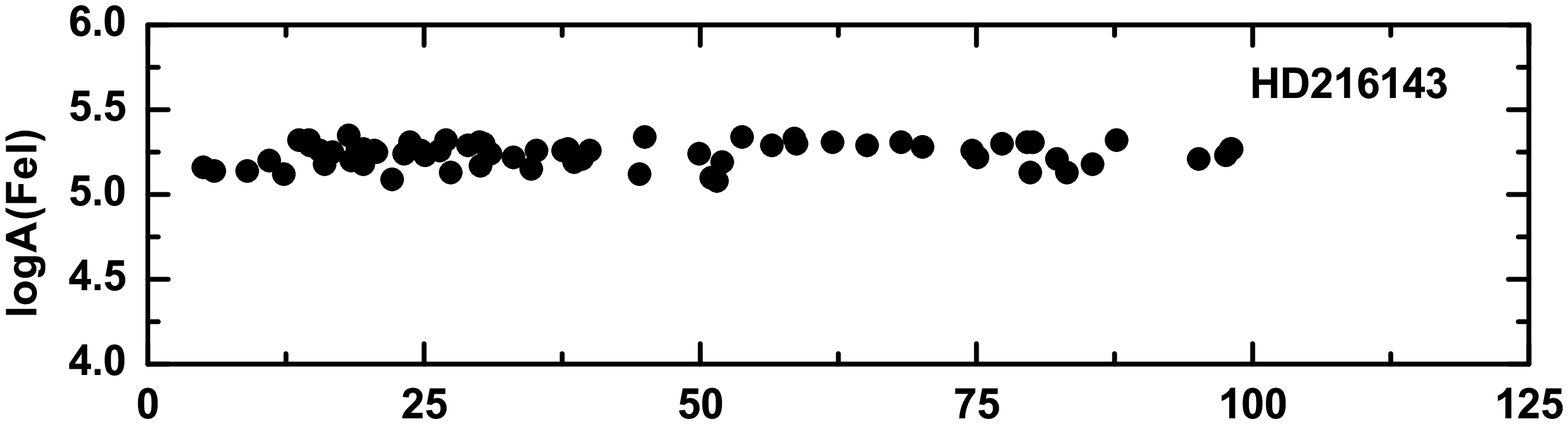}\\
\includegraphics[width=8.4cm]{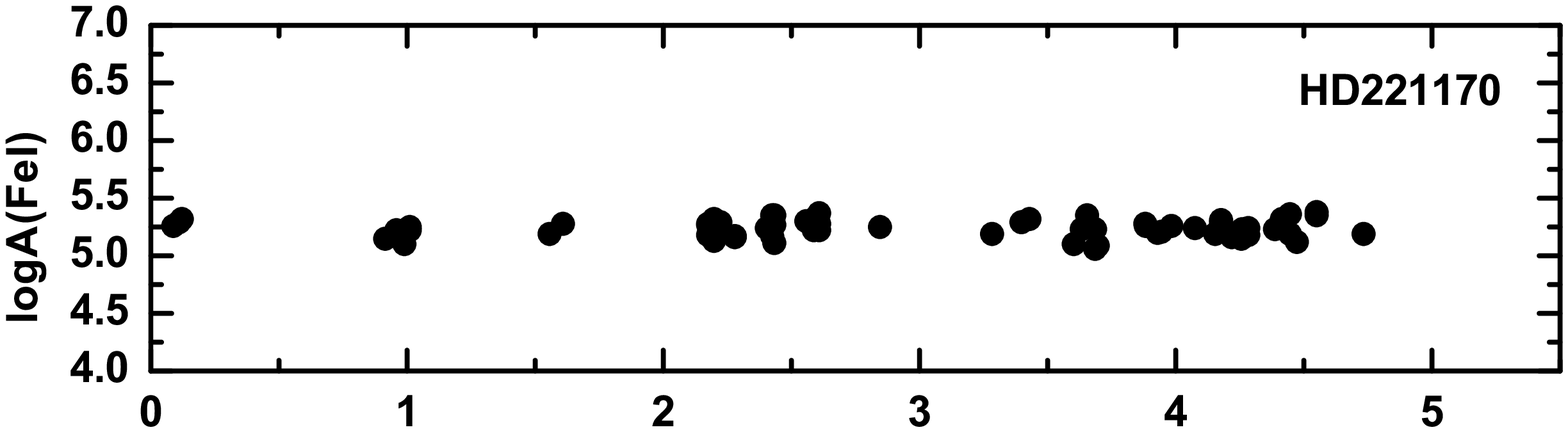}&\includegraphics[width=8.4cm]{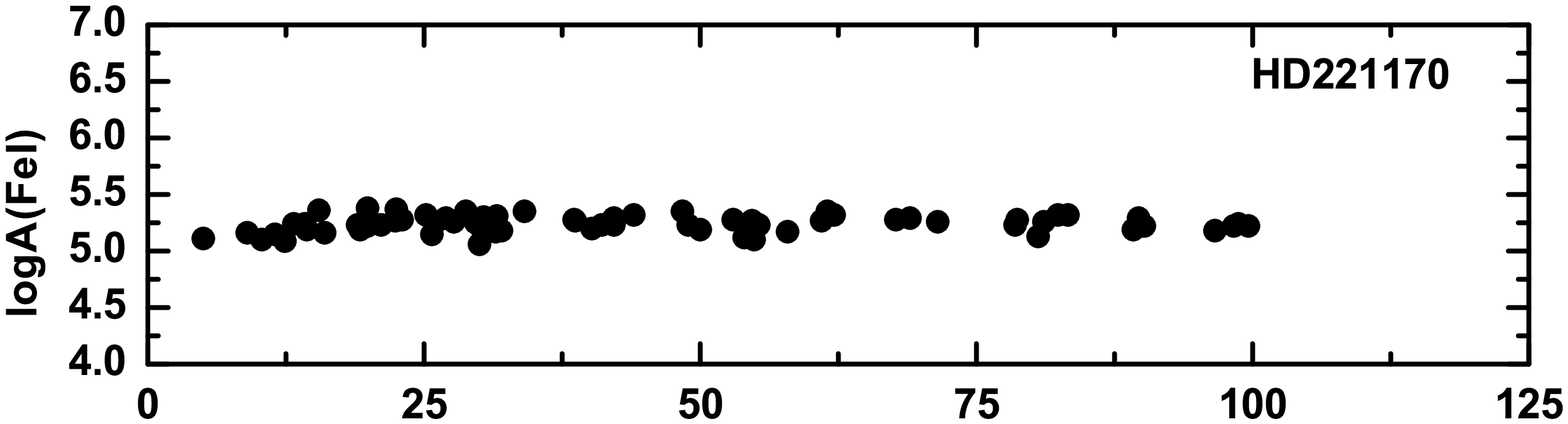}\\
\includegraphics[width=8.4cm]{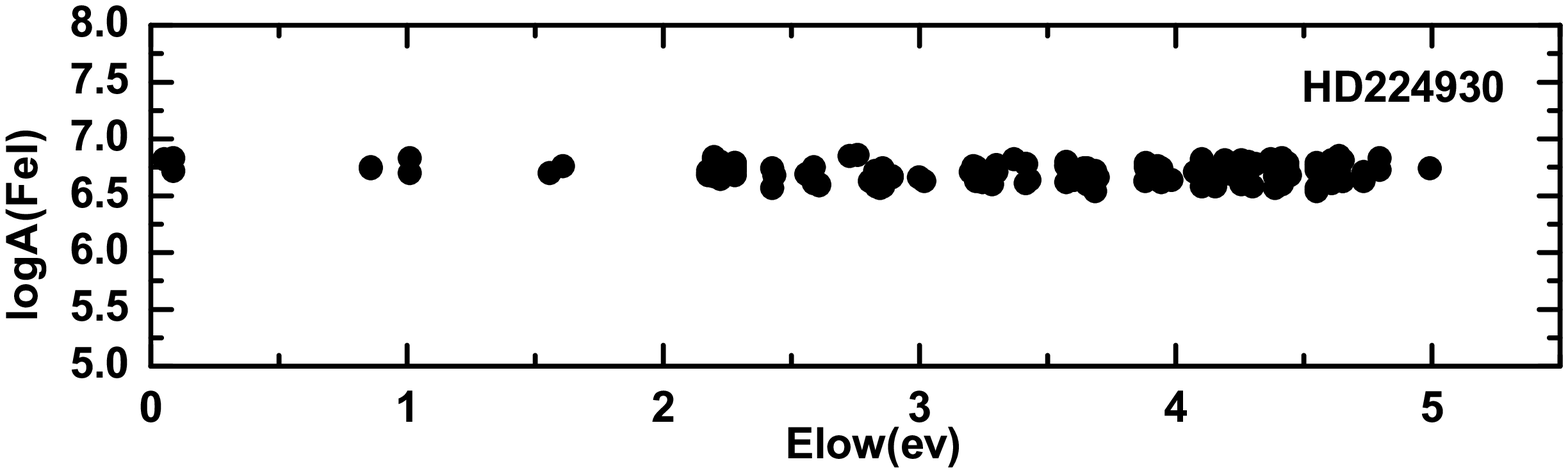}&\includegraphics[width=8.4cm]{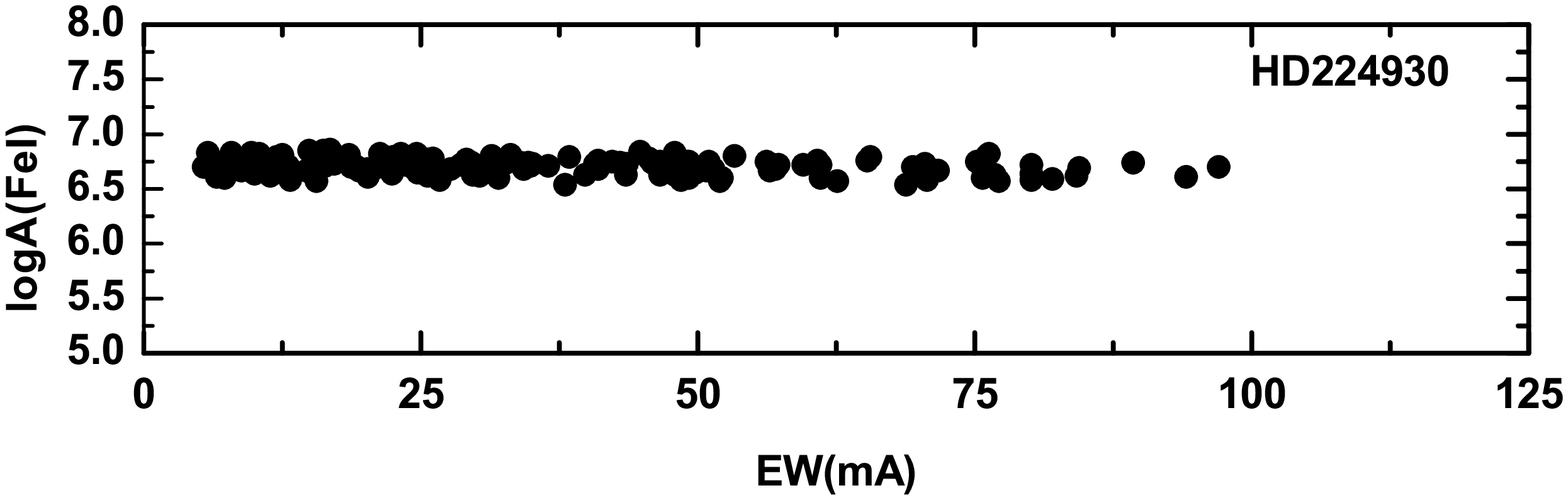}\\
\end{tabular}
\caption{(Fe/H) vs. Elow and EW.}
\label{elowew}
\end{figure*}

The turbulent velocity \Vt\ was defined by assuming the absence of 
correlation between the Fe abundance, estimated by the Fe~{\sc i} 
line, and its equivalent width EW (Figure \ref{elowew}). 
In Table \ref{param}, we compare our \Vt\ determinations 
with the calculations by \cite{sitn:15}, obtained by using 
the parametric formula calibrated over a large number of 
stars: \\

\Vt$= -0.21+0.06 [Fe/H]+5.6 ($\Teff/10$^4)-0.43$ \logg \\

The \Vt\ determinations obtained by this formula \citep{sitn:15} are in good agreement with our determinations, with the exceptions of: HD6582, 103095, and 224930. For these stars, our determinations are lower by 0.3 -- 0.5 dex. 
The star HD224930 is also included in \cite{takeda:07}, where they report a \Vt\ = 0.1 km/s, that is 0.2 dex lower than our estimation and 0.8 dex lower than Sitnova's formula \citep{sitn:15}.

The adopted value of the metallicity [Fe/H] was calculated using the Fe abundance obtained from the Fe~{\sc i} lines.

\subsection{Comparison of our parameter values with other authors, and error determinations}

In this section, we compare our atmospheric parameters to those obtained by
other authors and we estimate the impact of deviations from LTE on the parameter determination.

The stars HD6582, HD22879, HD84937, and HD103095 are Gaia benchmark stars which have \Teff\ and \logg\ determined from fundamental relations,
independently of spectroscopy \citep[][]{heiter:15}. 
These values and ours are compared in Table \ref{comp_heit}.

\begin{table*}
\begin{center}
\caption{Parameters of our target stars and comparison with \citet{heiter:15} for 4 common Gaia benchmark stars.}
\label{comp_heit}
\begin{tabular}{lccccccccc}
\hline
\multicolumn{1}{c}{} & \multicolumn{4}{c}{ \cite{heiter:15}} &$|$& \multicolumn{4}{c}{our}  \\
\hline
  HD (name)  & \Teff, K & $\sigma, \pm$ & \logg &  $\sigma, \pm$  &$|$&  \Teff, K & \logg & $\Delta$ \Teff, K& $\Delta$ \logg \\
\hline
 6582  ($\mu$ Cas) & 5308 & 29 &[4.41] &[0.06]&$|$& 5350& 4.5 &  42 &0.09 \\
 22879             & 5868 & 89 &  4.27 &  0.04&$|$ & 5825& 4.42& --43&0.15 \\ 
 84937             & 6356 & 97 &  4.06 &  0.04&$|$ & 6325&3.95 & --31&--0.11 \\
103095 (Gmb 1830)  &[4827]&[55]&  4.60 &  0.03&$|$ & 5100&4.65 & 273 &0.05\\
\hline                                                                                             
\end{tabular} 
\end{center}  
Note: The values between square brackets are not obtained directly, therefore we consider a two $\sigma$ error.
\end{table*}

We see agreement between our data and those by \cite{heiter:15} within the stated error definitions, except for the star HD103095 (Gmb 1830). 
However, there is no agreement between our and \cite{heiter:15} results, considering two $\sigma$ errors.
A detailed discussion that is useful to explain the temperature discrepancy for this star was presented in \cite{heiter:15}. The authors preferred the value \Teff\ obtained by the method, based on measuring the diameter of star. However, they summed up that the further interferometric observations at longer baselines and/or shorter wavelengths are clearly needed to resolve or confirm the \Teff\ discrepancy for Gmb 1830. The temperature values obtained for this star in other works are 5168 K \citep{cas:11} and 5129 K \citep{gonz:09},  are much closer to the value that we used.
 
The parameter determinations by other authors in the last fifteen years are presented in Table \ref{compar} for the same stars. 

\textbf{HD6582}. The average values of the parameters  obtained in  different studies 
are \Teff\  = 5336 K; \logg\ = 4.44; [Fe/H] = --0.86, and they agree with our determinations within one $\sigma$. 
The star ($\mu$ Cas) is one of Gaia FGK benchmark stars  \citep{jofre:14}.      

\textbf{HD6833}.  
The mean parameters for this star obtained in different studies  
are \Teff\ = 4425 K; \logg\ = 1.32; [Fe/H] = --0.95, and they agree with our determinations. 
This star has a peculiar chemical composition, and belongs to the CN-weak giants 
\citep[][]{luck:91}.

\textbf{HD19445}. This is a well known benchmark star for many studies.
The mean parameters for this star 
are \Teff\ = 5973 K; \logg\ = 4.34; [Fe/H] = --2.03. 
Our \Teff\ determinations differ from the mean value by 143 K. 
This is the only star for which the difference 
exceeds our assumed accuracy ($\pm$100 K), but they are still consistent within one $\sigma$.
High \Teff\ values are also reported by \cite{cas:10} and \cite{vanden:14}. 
In the study by \cite{cas:10}, \Teff\ was determined using the method based on the infrared fluxes. The same \Teff\ determination was adopted by \cite{vanden:14}.

\textbf{HD22879}. This star is well studied, used for different comparison of stellar parameters and chemical composition \citep[][]{jofre:14,sitn:15}. 
Our determinations are consistent with the mean parameter values, which are \Teff\  = 5853 K; \logg\ = 4.37; [Fe/H] = --0.83. For this star 
our stellar parameters under LTE approximation are in good concordance with the NLTE determinations reported in \citep{sitn:15} within the limits of the stated accuracy. 
This is due to the fact that the difference in the determinations for the line (Fe~{\sc i} -- Fe~{\sc ii}) in LTE is $-0.06\pm0.08$, and it is very close to the determination made under NLTE approximation which equals $-0.03\pm0.08$.

\textbf{HD84937}. This star was analysed several times by previous work \citep[e.g.,][]{jofre:14,sitn:15}. Our determinations agree with the mean parameters for this star, \Teff\ = 6353 K; \logg\ = 4.04; [Fe/H] = --2.09. In \cite{bensby:14}, the given \Teff\ is much higher compared to other authors. 
Our results under LTE approximation also agree with the NLTE determinations by \cite{sitn:15}, within the stated accuracy. However, the difference in the LTE and NLTE determinations for the (Fe~{\sc i} -- Fe~{\sc ii}) lines for this star is slightly higher than for HD22879: $-0.06\pm0.11$ and $0.0\pm0.12$, respectively. 

\textbf{HD103095}. This star (Gmb 1830) is one of the Gaia benchmark stars 
\citep{jofre:14}. Our mean parameters are \Teff\ = 5066 K, \logg\ = 4.61, [Fe/H] = --1.29. They agree with the determinations under NLTE approximation by \cite{sitn:15}. 
The difference in the determinations under LTE and NLTE approximation is close to zero. 

\textbf{HD170153}. Our parameters determinations are in good agreement with 
the mean values \Teff\ = 6104 K, \logg\ = 4.25, [Fe/H] = --0.62.

\textbf{HD216143}. The differences between our results and the mean parameter values \Teff\ = 4526 K, \logg\ = 1.02, [Fe/H] = --2.20 are within the given errors. 

\textbf{HD221170}. The mean parameters are \Teff\ = 4481 K, \logg\ = 0.97, [Fe/H] = --2.12, in agreements with our results.
The star is one of the most well-known $r$-process stars, used as a benchmark for $r$-process nucleosynthesis in the early Galaxy and in comparison with the $r$-process  abundances in the solar system \citep[e.g.][]{fulb:00, burris:00, ivans:06, molenda:13}. 

\textbf{HD224930}. Our determinations are in good agreement with the mean parameters \Teff\ = 5429 K, \logg\ = 4.36, [Fe/H] = --0.78.
\par
The discrepancies between our results and the average values for the ten stars in our sample are given by:\\  
$<$(\Teff$_{our}$ $-$ $<$\Teff$_{star}$$>$)$>$ = $<\Delta$ \Teff$>$, K,\\
$<$(\logg$_{our}$ $-$ $<$\logg$_{star}>$)$>$ = $<$$\Delta$\logg$>$, \\
$<$([Fe/H]$_{our}$ $-$ $<$[Fe/H]$_{star}>$)$>$ = $<$$\Delta$[Fe/H]$>$\\ 
and are presented in Table \ref{par_liter}. The $\Delta$ values are the mean difference of our values minus the average values obtained in other studies.

\begin{table*}
\caption{The comparison of our parameter determinations  with those of other authors:
 mean differences and rms deviations, n - number of common stars.}
\label{par_liter}
\begin{tabular}{lcccc}
\hline          
  $<\Delta$ \Teff$>$, K & $<\Delta$ \logg$>$   & $<\Delta$[Fe/H]$>$  & n &        references      \\
\hline                                                       \\
$60 \pm166$    &   $0.04 \pm0.11$ &  - &4  &       \cite{heiter:15}  \\
$-11 \pm46$   &    $0.04 \pm0.14$&  - &3   &     \cite{heiter:15}        \\
$22 \pm96$   &   $-0.05 \pm0.20$ & $-0.09 \pm0.08$ &6 &     \cite{gratton:03}    \\
$29 \pm107$     &   $0.07 \pm0.15$   & $0.05 \pm0.14$ & 9&    \cite{fulb:00}      \\
$-16 \pm66$    &  $0.00 \pm0.14$  &  $-0.04 \pm0.10$ & 10 &   mean values    \\
\hline
\end{tabular}                                         
\end{table*}

\par 
Summing up, the stellar parameters derived in this work are in good agreement with the results obtained in the literature. 
Based on Table \ref{par_liter}, we derive as errors for the effective temperature 
$\Delta$\Teff = $\pm100~K$, for the surface gravity $\Delta$\logg = $\pm0.2$, and for the micro-turbulent velocity $\Delta$\Vt = $\pm0.1$.

\section{Ages and kinematic parameters}
\label{sec: ages, kinematics}

In \cite{holm:09}, \cite{takeda:07}, \cite{maldon:12}, \cite{vanden:14}, \cite{delgado:14},  \cite{bensby:14}, \cite{ramirez:12}, \cite{ramirez:13} the 
stellar ages 
were determined for seven stars included in our sample (see Table \ref{age}). 
The age spread does not usually exceed 2 Gyr. This is consistent with the stated accuracy across all studies except \cite{holm:09}. In their work, for HD19445 and HD22849 the authors provide a different value for \Teff\ for age determination.
A detailed study of the stellar ages for the stars HD19445 and HD84937 is presented by \cite{vanden:14}, using both evolution tracks and the Wilkinson Microwave Anisotropy Probe (WMAP) observations. 
In the study by \cite{ivans:06}, the HD221170 star age was estimated using the Th/Eu ratio and equals 11.7$\pm$2.8 Gyr. This result is consistent with cosmic age determinations by the WMAP experiment (14.1 Gyr, \cite{tegmark:04}, and 13.7 Gyr, \cite{spergel:03}), with determinations of the main-sequence turnoff ages for globular clusters (14 Gyr, \cite{cho:16}) and with the results by PLANCK \citep[13.80 $\pm$ 0.04 Gyr][]{planck:16}. 

In this work, we defined the age using the tool available online at http://stev.oapd.inaf.it/cgi-bin/param, using the stellar tracks by \cite{bressan:12} and \cite{girardi:02}. For our stars with [Fe/H] $<$ $-$2.10, we used as input parameter the same value of [Fe/H] = $-$2.10 for each from these stars because the tracks limited by [Fe/H] = $-$2.20.
The results of age determinations are presented in Table \ref{age}. For stars HD103095 and 216143 we found the differences that are more than $\pm$2 Gyr. However, the errors are large enough to allow for this: for HD103095 those are  7.075 $\pm$ 3.930 Gyr \citep{bressan:12} and 5.261 $\pm$ 4.089 Gyr \citep{girardi:02} and for 
HD216143 those are 6.516 $\pm$ 4.441 Gyr \citep{bressan:12} and 2.906 $\pm$ 2.938 Gyr \citep{girardi:02}.
So, we find an agreement with the data of other authors, within the determination accuracy.

\begin{table*}
\begin{center}
\caption{Ages of our target stars and comparison with data of other authors.}
\label{age}
\begin{tabular}{lcccccccccccc}
\hline
\hline
HD      &       Age (Gyr) & &    &     &     &    &     &   &   &  &  &    \\
        &  Bressan(2012)& Girardi (2002)           & 1   &   2  &  3   &  4  &  5   &   6& 7  &   8 &  9& 10 \\ 
\hline
6582  &11.10 & 9.31 &-  &10.19&   - &   -  &    - &  -      &     -   &    -  & 2.10 &  1.90 \\ 
6833  &8.96  &9.61  &-- &--   &   - &   -  &     -&      -  &      -  &      -&      &   \\         
19445 & 11.16 & 10.79&4.5       &--&     -&     12.5&   -&      -&      11.65&  13.22 &13.80&    13.50    \\      
22879 & 11.48& 11.74 &7.5       &--&    -&       -&     -&      -&      12.85&  13.02&  13.80&  13.30  \\         
84937 & 11.46&11.47 &-- &--&     -&     12.09&  11.38&  10.2&   --&     --& &  \\         
103095& 7.08&  5.26 &-- &10.19&  -&     -&      -&      --&     --&     13.87 &  &  \\ 
170153& 8.79& 9.97&5.3  &--&     -&     -&      -&      -   &    7.71&  6.93 &   &        \\      
216143& 6.52 & 2.91 &-- &--&     -&     -&      -&      --&     --&     --&      \\ 
221170&9.24 &9.55  &--  &--&     -&     -&      -&      -&      --&     -- &  \\          
224930& 10.83 &10.14 &14.7      &10.19& 12.7&   -&      -&       -  &    6.3&   14.46&   &              \\        
\hline
\hline                                                                                             
\end{tabular} 
\end{center}  
Notes: 1- \cite{holm:09}, 2 - \cite{takeda:07}, 3 -  \cite{maldon:12}, 
4 -\cite{vanden:14}, 5 - \cite{delgado:14},  6 - \cite{bensby:14}, 
        7 - \cite{ramirez:12}, 8 - \cite{ramirez:13}, 9,10 - correspond to Padova and BASRI 
        isochrones, see in detail in \cite{cas:11}.                                                              
\end{table*}

\par
Kinematic and spatial characteristics of our target stars are important features in terms of their location in the Galaxy and dynamic evolution of the Galaxy. 
Distances and heliocentric velocities have been derived with the parallaxes and proper motions from the newly released TGAS catalogues \citep{tgas} or from Hipparcos \citep{leeuw:07}, combined to the radial velocities of Table 1. The orbital parameters have been computed by integrating the equations of motion in the galactic model of \cite{allen:93}, adopting a default value of 10 Gyr as the integration time. The adopted velocity of the Sun with respect to the LSR is (9.7, 5.2, 6.7) km/s 
\citep{bienayme:99}, the solar galactocentric distance R = 8.5 kpc and circular velocity V$_{LSR}$=220 km/s. The kinematical parameters are reported in Table \ref{kinpar}.

\begin{table*}
\caption{Kinematical parameters of studied stars.}    
\label{kinpar}
\begin{tabular}{lclccccccccccccl}         
\hline                 
HD   &		Dist & Ref	& err$_{\pi}$ &	U$_{v}$ &	V$_{v}$ &	W$_{v}$ &	 ecc &	Rmin &	Rmax &	Rmean &  Zmin &  Zmax & Zmean & Pop \\                         
     &	 	pc&   ($\pi$) &	\%&	                  km/s &     km/s&	km/s&		 &	kpc&	kpc &	kpc&  kpc&  kpc&  kpc&   \\                          
\hline                
6582    &     8&H& 0.5&  --44&--156&--36& 0.71& 1.5&  8.6&  6.01& -0.4& 0.4& 0.2	&halo or thick \\       
6833    &   197&T&10.0&   123&--202&  68& 0.93& 0.4& 10.9&  7.1& -6.5& 6.6& 1.9	&accreted halo\\       
19445   &    39&T& 0.9&   157&--123& -68& 0.69& 2.3& 12.2&  8.75& -1.5& 1.5& 0.8	&halo   \\       
22879   &    26&T& 0.8& --120& --81&--38& 0.47& 3.7& 10.4&  7.7& -0.5& 0.5& 0.3	&halo or thick \\       
84937   &    80&H& 8.5&   226&--238& --7& 0.98& 0.2& 15.4& 10.6& -9.2& 9.1& 1.6	&halo    \\       
103095  &     9&H& 0.7&   281&--158& -13& 0.91& 1.1& 22.1& 15.6& -0.2& 0.2& 0.1	&halo   \\       
170153  &     8&H& 0.4&     3&   44& --1&  0.23& 8.5& 13.5& 11.2& -0.1& 0.1& 0.1	&thick disc  \\     
216143  &  357 &T&20.4&   175&--168& 45 & 0.83& 1.2& 12.5&  8.8& -1.5& 1.5& 0.6 &halo    \\       
221170  &  448 &T&14.8&   105&--132&--23& 0.63& 2.3& 10.0&  7.2& -0.3& 0.4& 0.2 &halo   \\       
224930  &    12&H& 3.8&   --7& --76& -29&  0.32& 4.4&  8.5&  6.7& -0.3& 0.3& 0.2	&thick disc  \\       
\hline 
\end{tabular} 
Notes: Dist - distance; Ref - parallax $\pi$  sources:  H -  \citep{leeuw:07},  T -  \citep{tgas};  
 err$_{\pi}$ - parallax determination error; 
U$_{v}$, V$_{v}$, W$_{v}$ - components of heliocentric space velocity; ecc - eccentricity; 
Rmin, Rmax, Rmean - galactic distances; Zmin, Zmax, Zmean - 
distances from the galactic plane;  Pop - belonging to the type of the galaxy population.
\end{table*}     

\section{Selection of lines}
\label{sec: select lines}

Since the stars in our sample cover a large range of temperatures (4400K $<$ \Teff $<$ 6300K, including F, G, K-type dwarfs and giants) and metallicities ($-$2.26 $<$ [Fe/H] $<$ $-$0.6), it is necessary to create a list (or lists) of unblended lines for different parameter ranges to determine the chemical composition of the stars.
To create the list for iron and $s$-and $r$-process capture lines, we adopted data from several studies, including \cite{sneden:09}; \cite{lawler:06}, \cite{lawler:09}; \cite{takeda:05}; \cite{lai:08}; \cite{aoki:07}; \cite{ramirez:07}; \cite{simm:04}; \cite{den:06}; \cite{coluc:12},  \cite{bensby:14}, \cite{roed:14}. 
We selected about four thousand lines of different chemical elements. The atomic parameters of the list of lines were taken from the VALD database \citep{kupka:99}.
The Fe~{\sc i}, Fe~{\sc ii}, Ti~{\sc i}, Ni~{\sc i}, Cr~{\sc i}, Co~{\sc i} and V~{\sc i} lines were selected using the high-resolution solar spectrum obtained with the same spectrograph SOPHIE. Amongst more than 1000 initially selected lines, we choose from 102 (HD84937) to 303 (HD6582) unblended lines, which are optimal for the measurements at different metallicities.
Nine of the Mn~{\sc i} (4502, 4709, 4761, 4762, 4783, 4823, 5432, 6013, 6021 \AA\AA) lines were used to determine the manganese abundance; this list of lines was reported in the study by Prochaska \& McWilliam \citep{prochmcw:00} .
The abundances of neutron-capture elements (Y, Zr, La, Pr, Nd, Sm, Eu and Gd) were determined by the lines Y~{\sc ii} (4--10 lines), Zr~{\sc ii} (2--10 lines), La~{\sc ii} (4--9 lines), Pr~{\sc ii} (1--5 lines), Nd~{\sc ii} (5--11 lines), Sm~{\sc ii} (9--11 lines), Eu~{\sc ii} (1--2 lines) and Gd~{\sc ii} (3--5 lines). 
The list of the lines with their atomic parameters and equivalent widths in the spectra of all stars is presented in Table \ref{EWline}, the atomic data for the lines used in synthetic method 
calculations are presented in Table A1.

\begin{table}
\caption{Atomic data and equivalent widths EW of used lines.}
\label{EWline}
\begin{center}  
\begin{tabular}{lccccc}
\hline
HD &  $\lambda$,  & El &   EW & log gf  &   E$_{low}$,  \\
   & \AA\   & &   m\AA\~       &               &        eV    \\
\hline
6582 &   5517.533&   Si~{\sc i} &    6.7   &  --2.609 & 5.082  \\
6582 &   5645.613&   Si~{\sc i} &    16.2  &  --2.139 & 4.930  \\
6582 &   5665.554&   Si~{\sc i} &    18.1  &  --2.039 & 4.920  \\
6582 &   5684.484&   Si~{\sc i} &    33.0  &  --1.649 & 4.954  \\
 --  &    --     &   --  &    --    &  --     &  --    \\      
\hline                                                 
\end{tabular} 
\end{center}  
Notes:Table 7 are only available in electronic form
\end{table}

\section{Determination of chemical compositions (elemental abundances)}
\label{sec: abundance determination}

\begin{figure}
\begin{tabular}{c}
\includegraphics[width=8cm]{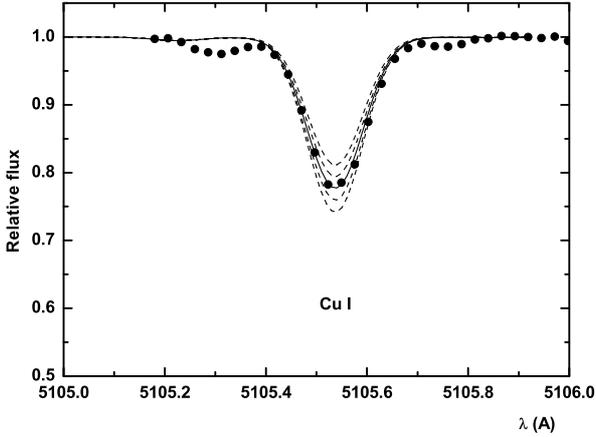}\\
\end{tabular}
\caption{Observed (dots) and calculated (solid and dashed lines) spectra in the region 
of Cu I line  for HD22879, the change in the Cu abundance is 0.05 dex.}
\label{pr_Cu}
\end{figure}

\begin{figure}
\begin{tabular}{c}
\includegraphics[width=8cm]{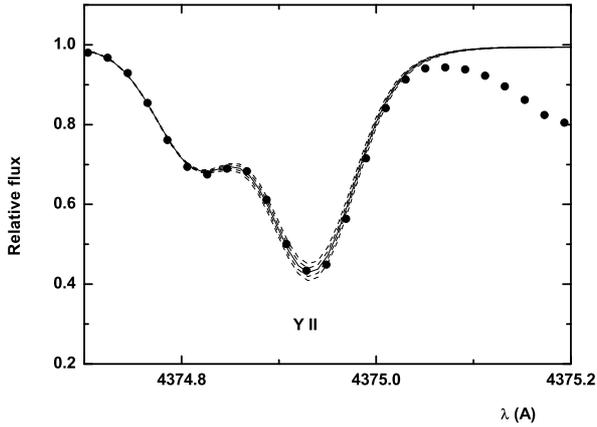}\\
\end{tabular}
\caption{Observed (dots) and calculated (solid and dashed lines) spectra in the region 
of Y II line  for HD22879, the change in the Y abundance is 0.05 dex.}
\label{pr_Y}
\end{figure}

The abundances of the investigated elements: Li, O, Na, Mg, Al, Si, Ca, Ni, Co, Mn,  Y, Zr, Ba, La, Ce, Nd, Sm, Eu, and Gd were determined for our target stars under LTE and NLTE approximations using the atmosphere models by \cite{castelli:04}. 
The Fe, Ti, V, Cr, Ni and Co abundances were determined using the equivalent widths EWs and the WIDTH9 code by \cite{kurucz:93}.
The Mn, Y, Zr, La, Pr, Ce, Nd, Sm, Eu and Gd abundances were determined using a new version of the STARSP synthetic spectrum code \citep{tsymbal:96}. The oscillator strengths for these lines were adopted from the VALD database \citep{kupka:99}. The Mn, Eu and Pr abundances were estimated accounting for the hyperfine structure: for the Mn~{\sc i} lines the HFS data were taken from \cite{prochmcw:00}. The van der Waals damping constant C6 was adopted from \cite{berg:08}. The HFS data for the Eu~{\sc ii} lines (4129 and 6645 A) were adopted from \cite{ivans:06}, and for the Pr~{\sc ii} lines from \cite{sneden:09}. The La and Sm lines were so weak that the hyperfine splitting (HFS) can be neglected. 
The spectrum synthesis fitting of the Cu and Y lines to the observed profiles is shown in Figs. \ref{pr_Cu},\ref{pr_Y}.

The abundances of the investigated elements were determined by differential analysis relative to the solar abundances. Solar abundances were calculated using the solar EWs, measured in the spectra of the Moon and asteroids; they were also estimated using the SOPHIE spectrograph and the oscillator strengths log\,gf adopted from the VALD database \citep{kupka:99}. 
The adopted solar abundance is presented in Table \ref{solar}.

\begin{table}
\caption{Used Solar abundance derived by us and other authors and compared with 
phosphoric abundance by \citet{asplund:09}.}
\label{solar}
\begin{tabular}{lll}
\hline
Species &    log A (our definitions+)  &    Asplund et al. (2009)   \\
 \hline
 Li I    &    1.10   (1)            &  1.05 $\pm$0.10         \\ 
 O I     &    8.71 $\pm$0.05 (7)    & 8.69 $\pm$0.05  \\
Na I     &    6.25 $\pm$0.04 (8)   &  6.24 $\pm$0.04  \\
Mg I     &    7.54 $\pm$0.03 (18)   &  7.60 $\pm$0.04  \\
Al I     &    6.43 $\pm$0.04 (9)    &  6.45 $\pm$0.03  \\
Si I    &    7.52 $\pm$0.08 (18)   &  7.51 $\pm$0.03  \\
Ca I    &    6.31 $\pm$0.04 (45)   &  6.34 $\pm$0.04  \\
Sc II   &    3.06 $\pm$0.03 (8)   &   3.15 $\pm$0.04   \\
Ti I    &    4.92 $\pm$0.05 (23)   &  4.95 $\pm$0.05  \\
Ti II   &    5.01 $\pm$0.01 (18)    &        --         \\
$<$ Ti $>$&  4.97     &4.95 $\pm$0.05  \\
V I    &    3.96 $\pm$0.06 (25)   &  3.93 $\pm$0.08  \\
Cr I    &    5.56 $\pm$0.08 (20)   &  5.64 $\pm$0.04  \\
Cr II   &    5.74 $\pm$0.10 (7)   &  5.64 $\pm$0.04  \\
$<$ Cr $>$    &     5.65                &5.64 $\pm$0.04    \\
Mn I    &    5.22 $\pm$0.08 (11)   &  5.43 $\pm$0.05  \\
Fe I    &    7.50 $\pm$0.10 (140)  &  7.50 $\pm$0.04  \\
Fe II   &    7.50 $\pm$0.11 (13)   &        --         \\
Co I    &    4.96 $\pm$0.10 (15)   &  4.99 $\pm$0.07  \\
Ni I    &    6.20 $\pm$0.07  (32)  &  6.22 $\pm$0.04  \\
Cu I    &   4.06 $\pm$0.04  (3)  &  4.19 $\pm$0.04 \\
Zn I    &    4.54 $\pm$0.05  (3)  &  4.56 $\pm$0.05  \\
Sr II   &    2.92 $\pm$0.07  (5)  &  2.87 $\pm$0.07  \\
Y II    &    2.07 $\pm$0.05  (5)   &  2.21 $\pm$0.05  \\
Zr II   &    2.60  $\pm$0.03 (3)   &  2.58 $\pm$0.04   \\
Ba II   &    2.17 $\pm$0.05 (4)    &  2.18$\pm$0.09   \\
La II   &    1.10 $\pm$0.06 (5)    &  1.10 $\pm$0.04  \\
Pr II   &    0.81 $\pm$0.00 (2)         &   0.72 $\pm$0.04     \\
Nd II   &    1.46 $\pm$0.05  (4)  &  1.42 $\pm$0.04  \\
Sm II   &    1.00 \citep{lawler:06}   &  0.96 $\pm$0.04  \\
Eu II   &     0.42   (4129)      &  0.52 $\pm$0.04  \\
Eu II   &     0.51  (6645)      &  0.52 $\pm$0.04  \\
Gd I   &     1.11  \citep{den:06}      &  1.07 $\pm$0.04  \\
\hline
\end{tabular}
\end{table}

\subsection{Lithium abundance}
The Li abundances in the investigated stars were obtained by fitting the observational profiles to the synthetic spectra that were computed by the STARSP LTE spectral synthesis code, developed by \cite{tsymbal:96}. Considering the wide temperature and metallicity ranges of the target stars, we used the detailed list of the atomic and molecular lines in the region of the $^7${Li} 6707 \AA\ line  \citep{mish:97}.
The comparison of the Li abundance determinations with the results obtained by other authors is given in Table \ref{li}. We can see from this Table that our Li determinations are in good agreement with ones of the others authors.

\begin{table}
\caption{Lithium abundance.}
\label{li}
\begin{tabular}{lcccccc}
\hline
  HD  &  logA(Li)&-&-&-&-\\
	    &    our &  up${_(our)}$ &  1   &  up${_1}$  &  2 & 3\\
\hline	
6582	&	--	&0.20	&--	&0.4  &  -- & -  \\
6833	&	--	&-0.20	&--	&--&	--  & -  \\
19445	&	2.10	&--	&2.26	&--&	2.10  & 2.25\\
22879	&	1.50	&--	&1.59	&--&	 --  & 1.44 \\
84937	&	2.31	&--	&2.31	&--&	 --   & 2.4  \\
103095	&	0.51	&--	&--	&--&	 --  & 0.42\\
170153	&	2.37	&--	&2.41	&--&	 -- & -  \\
216143	&	--	&-0.50	&--	&--&	 -- & -  \\
221170	&	--	&-0.40	&--	&--&	 -- & -  \\
224930	&	--	&0.30	&--	&--0.51& -- & -  \\
\hline 
\hline
\end{tabular} 
Notes: up${_(our)}$ - the upper limit of Li abundance,
1  - Li abundance and  up${_1}$  - the upper limit of Li abundance \citep{ramirez:12};
2 - \cite{roed:14};
3 - \cite{fulb:00}.
\end{table}

\subsection{NLTE abundance determinations}

\begin{figure*}
\includegraphics[width=15.4cm]{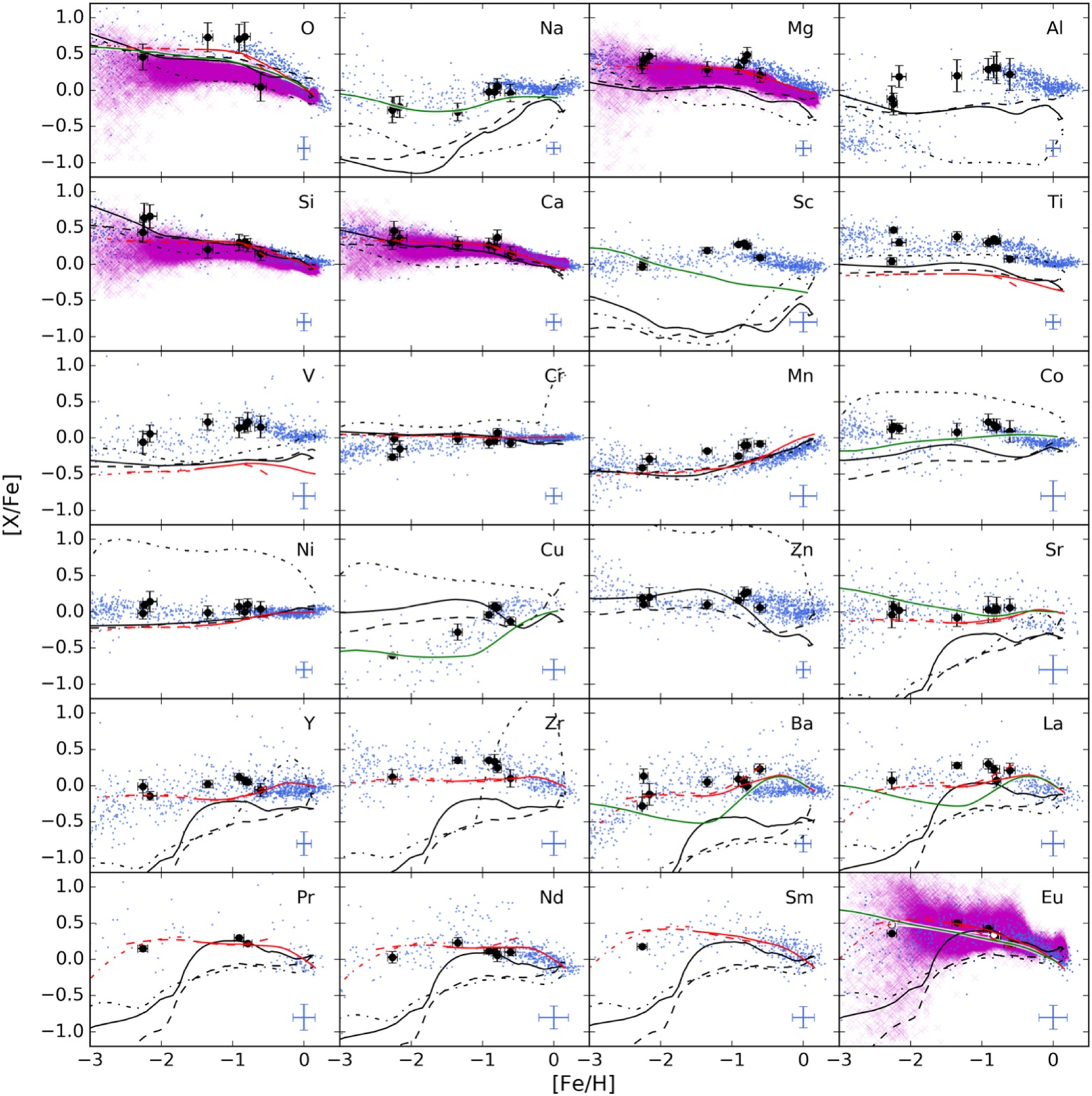}
\caption{The elemental abundances [X/Fe] with respect to [Fe/H] is shown for our target stars
, in comparison with a large sample of stellar observations in the galaxy, and with predictions from different galactic chemical
evolution codes. Our elemental abundances were presented as points with error bars corresponding to the results fin Table \ref{abund} (note, not all stars have a determined
abundance of all investigated elements).
The observation data from the literature are marked as blue dots \protect\cite{aoki:08,roederer:09,roederer:14,hansen:12,ishigaki:12,ishigaki:13,cohen:13,bensby:14,battistini:15,battistini:16}.
Additional data have been used for Cu \protect\cite{reddy:06} and Pr \protect\cite{hinkel:14}.
Black lines have been produced by OMEGA, a one-zone model (solid and dashed lines correspond to the massive star yields of West \& Heger (private communication) 
using the \protect\cite{ertl16} and the \textit{no-cutoff} prescriptions for the stellar remnant masses, respectively.
The black dotted lines represent  NuGrid Set 1 extension massive star yields. The GCE model predictions by \protect\cite{bisterzo:14} are shown
with red lines (solid line - thin disk, dashed line - thick disk, dashed-dotted line - halo).
The green solid line corresponds to the model predictions associated with the solar neighbourhood chemical evolution model described by
\protect\cite{hughes:08}, realised with the GEtool software package. Results from the ICE code \protect\cite{Wehmeyer15} are shown with magenta crosses.}

\label{abfig}
\end{figure*}

The Na, Mg, Al, Ca, Sr and Ba abundances were computed under NLTE approximation 
with a version of MULTI \citep{carl:86}, modified by S. Korotin \citep{kor:99}. 
The model of Na atom consists of 27 levels of Na I and the ground level of 
Na~{\sc i}. We considered the radiative transitions between the first 20 levels of 
Na~{\sc i} and the ground level of Na~{\sc ii}. Transitions between the remaining levels were used only in the equations of particle number conservation. In the linearisation procedure, 46 b-b and 20 b-f transitions were included.     
We employed the model of Mg atom consisting of 97 levels: 84 levels of 
Mg~{\sc i}, 12 levels of Mg~{\sc ii} and a ground state of Mg~{\sc iii}. Within the described 
system of the Mg atom levels, we considered the radiative transitions 
between the first 59 levels of Mg~{\sc i} and ground level of Mg~{\sc ii}. Transitions 
between the rest levels were not taken into account and were used only in the 
equations of particle number conservation. A more detailed description of these 
computations can be found in \cite{mish:04}.         
To derive the NLTE Al abundances, we used the Al~{\sc i} lines at 3944, 3961, 
5557.06, 6696.02, and 6698.67 \AA\AA. Our Al atomic model is described in detail 
in \cite{andr:08}. 
This model atom consists of 76 levels of Al~{\sc i} and 13 levels
of Al~{\sc ii}.
Our model of Ca atom consists of 70 levels of Ca~{\sc i}, 38 levels of Ca~{\sc ii}, 
and the ground state of Ca~{\sc iii} were taken into account; in addition, more than 
300 levels of Ca~{\sc i} and Ca~{\sc ii} were included to keep the condition of the 
particle number conservation in LTE. The information about the adopted 
oscillator strengths, photoionisation cross-sections, collisional rates and 
broadening parameters can be found in \cite{spite:12}. For the analysis we use 45 Ca lines in the visible spectrum.    
In our analysis a Sr ion model includes 44 low levels of Sr~{\sc ii} with 
n $\lid$ 12 and l $\lid$ 4 and the ground level of Sr~{\sc iii}. It also accounts for 
the fine splitting under the terms 4d2D and 5p2P$^{0}$. That is why we included 24 
Sr~{\sc i} levels only into the equation of particle number conservation. A more 
detailed description of the model atom can be found in \cite{andr:11}. 
The lines from the blue section of the spectrum, such as resonance lines 4077 and 
4215 \AA, as well as subordinate line 4161 \AA, were used in the present study.        
Our Ba model contains 31 levels of Ba~{\sc i}, 101 levels of Ba~{\sc ii} with $n < 50$, 
and the ground level of Ba~{\sc iii} ion. 91 bound-bound transitions between the 
first 28 levels of Ba~{\sc ii} ($n < 12$ and $l < 5$) were computed in detail. The 
odd Ba isotopes have hyperfine splitting of their levels and, thus, several 
Hyper Fine Structure (HFS) components for each line \citep{rut:78}. The 
information about the adopted oscillator strengths, photoionization 
cross-sections, collisional rates and broadening parameters can be found in  
\cite{andr:09}. 
The abundance determinations for the studied elements are given in Table \ref{abund} and are presented in Figure \ref{abfig}.

For comparison, stellar data observed in the Galaxy are shown from the following references: \cite{reddy:06, aoki:08,roederer:09,roederer:14,hansen:12,ishigaki:12,ishigaki:13,cohen:13, bensby:14, 
hinkel:14, battistini:15, battistini:16}. 
The astrophysical implications and discussion based on these observations are given in section \ref{sec: result, discuss}.

\begin{figure}
\begin{tabular}{c}
\includegraphics[width=8cm]{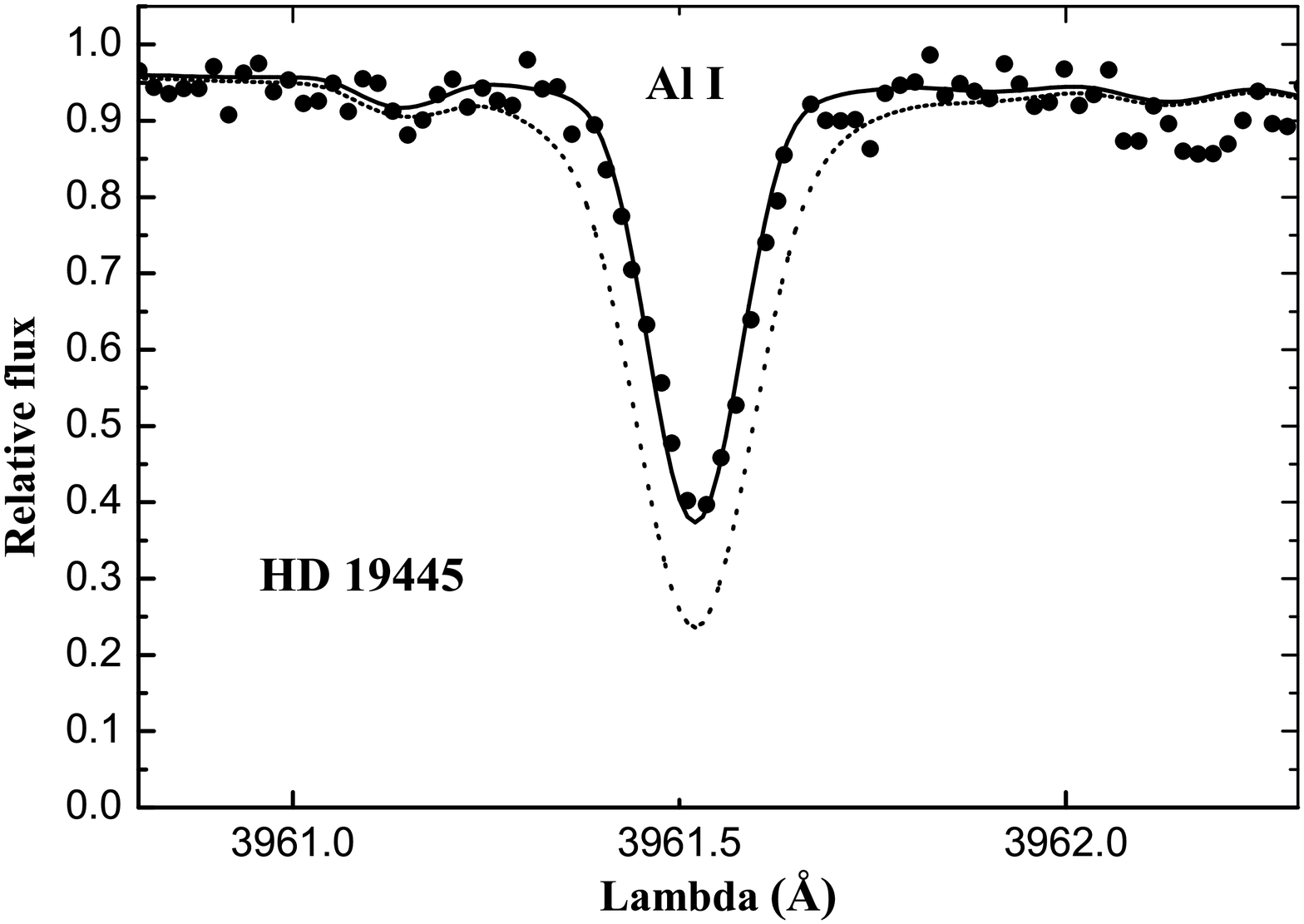}\\
\includegraphics[width=8cm]{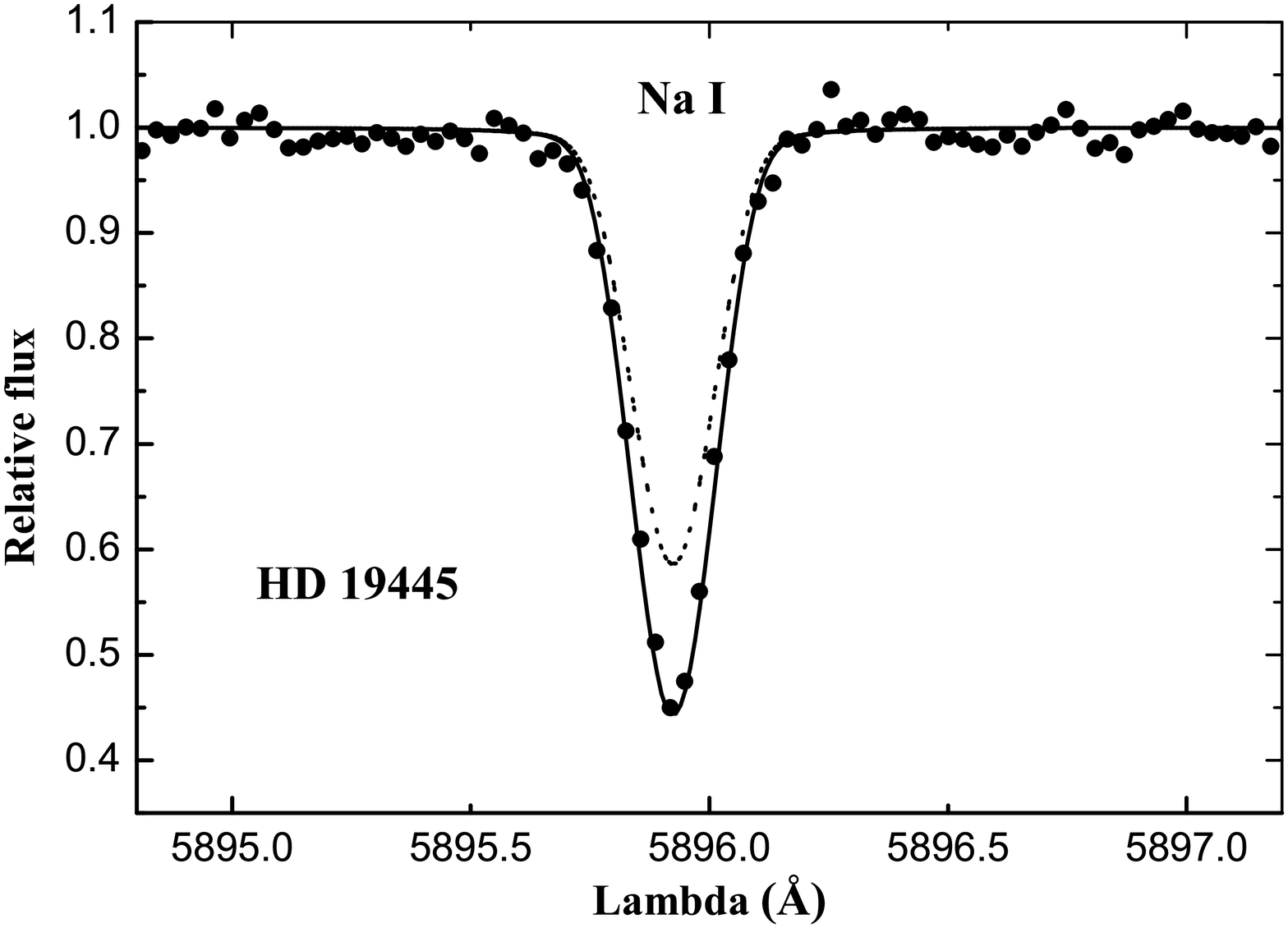}\\
\end{tabular}
\caption{Observed (dots) and calculated (solid - NLTE and dashed - LTE lines) spectra in the region 
of Al I and Na I lines for HD19445.}
\label{nlteline}
\end{figure}

\begin{table*}        
\caption{Elemental abundance of our target stars.}   
\label{abund}                                        
\begin{tabular}{rrrrrrrrrrr}                             
\hline                                                                                         		
HD		&6582	&6833	&19445	&22879	&84937	&103095	&170153	&216143	&221170	&224930	\\            
\hline													
[Fe/H]		&--0.83	  &--0.77	&--2.16	&--0.91	&--2.24	&--1.35	&--0.61	&--2.26	&--2.26	&--0.79  \\
$\sigma, \pm$	&0.08	&0.07	&0.10	&0.07	&0.06	&0.07	&0.08	&0.07	&0.06	&0.06	\\

 [O/Fe]		&0.74	&0.08	& 	&0.71	& 	&0.73	&0.05	&0.46	&0.46	&	\\
$\sigma, \pm$	&0.20	&0.10	& 	&0.20	&	&0.20	&0.20	&0.18	&0.18	&	\\
 
[Na/Fe]		&--0.02	&--0.37	&--0.23	&--0.02	&--0.23	&--0.30  &--0.04 &--0.27 &--0.27 &0.06	\\
$\sigma, \pm$	&0.12	&0.14	&0.15	&0.12	&0.14	&0.12	&0.12	&0.18	&0.18	&0.10	\\

[Mg/Fe]		&0.41	&0.04	&0.47	&0.32	&0.42	&0.29	&0.21	&0.33	&0.33	&0.49	\\
$\sigma, \pm$	&0.10	&0.10	&0.10	&0.10	&0.10	&0.10	&0.10	&0.10	&0.15	&0.10	\\

[Al/Fe]		&0.32	&--0.30	&0.19	&0.29	&--0.19	&0.20   &0.22	&--0.12	&--0.22	&0.31	\\
$\sigma, \pm$	&0.15	&0.15	&0.15	&0.15	&0.15	&0.22	&0.22	&0.18	&0.18	&0.22	\\

[Si/Fe]		&0.31   & 0.16  &  0.66 &   0.30&  0.64 &   0.20 &  0.14 &  0.44 & 0.56  &  0.26 \\
$\sigma, \pm$	&0.08   & 0.10  &  0.16 &   0.11&  0.20&  0.12 &   0.09&  0.14	& 0.11  &  0.09	\\    

[Ca/Fe]		&0.24	&0.32	&0.36	&0.26	&0.46	&0.28	&0.14	&0.30	&0.46	&0.37	\\		
$\sigma, \pm$   & 0.10	&0.12	&0.10	&0.10	&0.13	&0.10	&0.10	&0.10	&0.10	&0.10	\\

[Sc/Fe]	       &0.29	&--0.07	&	&0.27	&0.05	&0.19	&0.09	&--0.04	&0.10	&0.25	\\                    
$\sigma, \pm$	&0.03	&0.04	&	&0.02   &0.03	&0.04	&0.03	&0.04	&0.05	&0.04	\\

[Ti/Fe]		&0.35	&0.09	&0.30	&0.30	&0.47	&0.38	&0.07	&0.04	&0.14	&0.32	\\
$\sigma, \pm$	&0.03	&0.06	&0.05	&0.06	&0.03	&0.07	&0.04	&0.05	&0.05	&0.04	\\

[V/Fe]		&0.16	&0.01	&0.06	&0.14	&	&0.22	&0.15	&--0.06	&0.11	&0.22	\\
$\sigma, \pm$	&0.09	&0.12	&0.12	&0.14	&	&0.11	&0.15	&0.15	&0.15	&0.13	\\

[Cr/Fe]		&--0.04   &--0.04 &--0.15 &--0.06 &--0.01 &--0.01 &--0.07 &--0.26 &--0.30& 0.07	\\
$\sigma, \pm$	&0.05	&0.07	&0.11	&0.08	&0.08	&0.08	&0.06	&0.04	&0.03	&0.06	\\

[Mn/Fe]		&-0.10	&-0.38	&-0.29	&-0.25	&	&-0.18	&-0.08	&-0.41	&-0.35	&-0.10	\\
$\sigma, \pm$	&0.07	&0.03	&0.08	&0.03	&	&0.02	&0.04	&0.03	&0.01	&0.09	\\

[Co/Fe]		&0.18	&--0.15	&0.13	&0.22	&0.16	&0.08	&0.09	&0.12	&0.00	&0.15	\\
$\sigma, \pm$	&0.08	&0.11	&0.07	&0.11	&0.03	&0.12	&0.15	&0.13	&0.08	&0.11	\\

[Ni/Fe]		&0.00	&--0.12	&0.14	&0.07	&0.09	&--0.01	&0.04	&--0.02	&0.01	&0.10	\\
$\sigma, \pm$	&0.07	&0.07	&0.14	&0.10	&0.05	&0.09	&0.10	&0.08	&0.08	&0.08	\\

[Cu/Fe]	        &0.08	&--0.54	&	&--0.04	&	&-0.28	&-0.13	&-0.60	&-0.63	&0.06	\\   
$\sigma, \pm$	&0.05	&0.04	&	&0.04	&	&0.11	&0.06	&	&	&	\\   

[Zn/Fe]	        & 0.25	&--0.01	&0.20	&0.16	&0.10	&0.10	&0.06	&0.19	&0.22	&0.27	\\   
$\sigma, \pm$	&0.09	&0.06	&0.13	&0.04	&0.00	&0.06	&0.07	&0.05	&0.03	&0.03	\\ 

[Sr/Fe]		&-0.01	&--0.14 &0.02	&0.03	&0.07	&--0.08	&0.06	&--0.04	&--0.08	&0.05	\\
$\sigma, \pm$	&0.12	&0.15	&0.15	&0.18	&0.15	&0.12	&0.12	&0.18	&0.18	&0.15	\\

[Y/Fe]		&0.07	&--0.21	&-0.14	&0.12	&	&0.02	&--0.06	&-0.01	&--0.17	&0.05	\\
$\sigma, \pm$	&0.05	&0.09	&0.06	&0.05	&	&0.05	&0.09	&0.09	&0.09	&0.04	\\

[Zr/Fe]		&0.33	&0.07	&	&0.35	&	&0.35	&0.1	&0.12	&0.32	&0.25	\\
$\sigma, \pm$	&0.10	&0.07	&	&0.02	&	&0.04	&0.12	&0.10	&0.06	&0.07	\\

[Ba/Fe]		&0.05	&0.05	&--0.12	&0.09	&0.13	&0.05	&0.23	&--0.28	&--0.12	&--0.01	\\
$\sigma, \pm$	&0.07	&0.10	&0.15	&0.10	&0.10	&0.07	&0.07	&0.12	&0.10	&0.12	\\

[La/Fe]		&0.23	&0.10	&	&0.3	&	&0.28	&0.21	&0.07	&0.31	&0.07	\\
$\sigma, \pm$	&0.05	&0.06	&	&0.07	&	&0.04	&0.10	&0.12	&0.08	&0.10	\\

[Pr/Fe]		&	&0.02	&	&0.3	&	&	&	&0.15	&0.39	&0.22	\\
$\sigma, \pm$	&	&0.07	&	&	&	&	&	&0.04	&0.10	&0.03	\\

[Nd/Fe]		&0.11	&0.11	&	&0.12	&	&0.23	&0.1	&0.03	&0.28	&0.06	\\
$\sigma, \pm$	&0.06	&0.06	&	&0.05	&	&0.07	&0.05	&0.08	&0.05	&0.09	\\

[Sm/Fe]		&	&	&	&	&	&	&	&0.18	&0.45	&	\\
$\sigma, \pm$	&	&	&	&	&	&	&	&0.03	&0.03	&	\\

[Eu/Fe]$_{4129}$	&0.35	&0.31	&	&0.43	&	&0.51	&0.21	&0.36	&0.59	&0.29	\\

[Eu/Fe]$_{6645}$	&0.34	&0.44	&	&	&	&	&	&0.48	&0.71	&	\\

[Gd/Fe]		&	&	&	&	&	&	&	&0.18	&0.47	&	\\
$\sigma, \pm$	&	&	&	&	&	&	&	&0.05	&0.07	&	\\
\hline                                                                                           	
\end{tabular}                                                                                    	
\end{table*}

\subsection{Errors in abundance determinations}

As can be seen from Table \ref{abund}, the O abundance has
the  largest error, ranging between 0.10 and 0.2 dex,  this being due to the O weak lines  that we used. The best measured abundances of Cr, Fe, Mn, the errors are between 0.03 and 0.11 dex.
To determine the systematic errors in the elemental abundance resulting from 
uncertainties in the atmospheric parameter determinations, we derived the elemental
abundance of two stars with different stellar parameters, HD170153 (\Teff = 6170 K, \logg = 4.25, \Vt = 0.7, [Fe/H] = --0.61) and HD221170 (\Teff = 4415 K, \logg = 1.05, \Vt = 1.9, [Fe/H] = --2.26) for several models with modified parameters 
($\Delta$\Teff $= \pm100~K$, $\Delta$\logg $= \pm0.2$, $\Delta$\Vt $= \pm0.1$). 
The abundance variations with the modified parameters,  the fitting errors
for the computed and observed spectral line profiles (0.03 dex) and total error (tot+), are given in 
Table \ref{errors}. The maximum contribution to the error is introduced by \Teff 
when neutral atom lines are used for the abundance determination 
and by \logg, if the lines of ionised atoms are used. Total error due to parameter uncertainties and the measured of the spectra varies from 0.05 -- 0.11 dex for the hot and more metallicity  stars and  to 0.06 -- 0.18 dex for the cold  metal-poor stars, for [Fe/H] it is from 0.08 to 0.12 dex, respectively.

\begin{table*}
\caption{Abundance errors due to atmospheric parameter uncertainties as examples of stars 
with different values of stellar parameters: HD170153 (\Teff = 6170 K, \logg = 4.25, \Vt = 0.7 km/s, [Fe/H] = --0.61) and HD221170 (\Teff = 4415 K, \logg = 1.05, \Vt = 1.9 km/s, [Fe/H] = --2.26).}
\label{errors}
\begin{tabular}{llccccccccc}
\hline
& & HD170153  & & & & $|$&HD221170 & & & \\
 AN & El  & $\Delta$ \Teff+  & $\Delta$ \logg+ & $\Delta$ \Vt+ & tot+ &$|$&$\Delta$ \Teff+  & $\Delta$ \logg+ & $\Delta$ \Vt+ & tot+\\
\hline
11	&NaI	&0.04	&--0.02	&--0.01	&0.05&$|$	&0.08	&--0.02	&--0.03	&0.09			 \\  
12	&MgI	&0.04	&--0.03	&--0.01	&0.05&$|$	&0.07	&--0.03	&--0.05	&0.09			 \\ 
13	&AlI	&0.04	&0.05   &  0.00 &0.06&$|$ 	&0.07	&--0.02	&--0.02	&0.08			  \\ 
14	&SiI	&0.05	&0.04	&--0.01 &0.06&$|$ 	&0.04	&0.00  &--0.04	&0.06			    \\ 
20	&CaI	&0.07	&--0.04	&--0.03	&0.09&$|$	&0.11	&--0.04	&--0.07	&0.14			  \\ 
23.01   &ScII   &0.03	&0.08  & --0.01 &0.09&$|$   &0.02	&0.07  &--0.01&  0.07     \\
22	&TiI	&0.08	&--0.01	&--0.02	&0.08&$|$	&0.18	&--0.02	&--0.01	&0.18			  \\ 
22.01	&TiII	&0.03	&0.07	&--0.03	&0.08&$|$	&0.02	&0.06  &--0.01  &0.06			  \\  
23	&VI     &0.09	&0.00  &--0.02	&0.09&$|$  	&0.18	&--0.03	&--0.01	&0.18		  \\  
24	&CrI	&0.09	&--0.02	&--0.04	&0.10&$|$	&0.17	&--0.03	&--0.04	&0.18			   \\ 
24.01	&CrII	&0.00	&0.07	&--0.03 &0.08&$|$	&--0.03	&0.07	&--0.02	&0.08		 \\ 
25	&MnI	&0.06	&0.00  &-0.01	&0.06&$|$	&0.12	&--0.02	&0.00	  &0.12		  \\ 
26	&FeI	&0.07	&--0.02	&--0.03	&0.08&$|$   &0.12	&--0.02	&--0.02	&0.12			 \\ 
26.01	&FeII	&0.00	&0.06	&--0.04	&0.07&$|$	&--0.03  &0.07 	&--0.02	&0.08		  \\  
27	&CoI	&0.09	&--0.03	&--0.05	&0.11&$|$	&0.17	&--0.03	&--0.04	&0.18			 \\  
28	&NiI	&0.06	&0.00  &--0.01	&0.06&$|$	&0.10	&--0.01	&--0.01	&0.10			 \\ 
29	&CuI	&0.08	&0.00  	&--0.01	&0.08&$|$	&0.14	&--0.02	&--0.01	&0.14		  \\ 
30	&ZnI	&0.05	&0.02   &--0.04	&0.07&$|$	&0.01	&0.04 	&--0.02	&0.05		  \\ 
38	&SrII	&0.05	&0.02  &--0.02	&0.06&$|$	&0.04	&0.05  &--0.08	&0.10		 \\ 
39	&YII	&0.04	&0.06  &--0.03	&0.08&$|$	&0.03	&0.07  &--0.01	&0.08		  \\  
40	&ZrII	&0.04	&--0.02	&0.08   &0.09&$|$	&0.03	&0.07  &--0.02	&0.08		 \\  
56	&BaII	&0.07	&0.01  	&--0.09	&0.11&$|$	&0.06	&0.07  &--0.06	&0.11		 \\ 
57	&LaII	&0.05	&0.09 	&--0.02	&0.10&$|$	&0.06	&0.07  &0.00	  &0.09		  \\ 
59	&PrII	&0.04	&0.07  &0.00 	&0.08&$|$	&0.06	&0.07  &0.00	  &0.09			  \\ 
60	&NdII	&0.04	&0.07  &0.00    &0.08&$|$	&0.06	&0.07  &0.00 	&0.09			\\ 
62	&SmII	&0.04	&0.07  &0.00    &0.08&$|$	&0.06	&0.07  &0.00	  &0.09		  \\  
63	&EuII	&0.04   &0.08  &0.00    &0.09&$|$	&0.07	&0.08 &0.04	 &0.11		 \\  
64      &GdII   &       &      &        &    &$|$	&0.03  & 0.07  &0.00   & 0.09			\\
\hline                             
\end{tabular}
\end{table*}

The comparison of our abundance determinations (1) with those from \cite{jofre:15} (2) are presented in Table \ref{abun_jofr} for the stars using as GAIA benchmark.

\begin{table*}
\caption{The comparison of our abundance determinations (1) with those from \citet{jofre:15} (2), and the difference ([El/H]${_1}$ - [El/H]${_2}$) (3). }
\label{abun_jofr}
\begin{tabular}{lccccccccccccccc}
\hline
\multicolumn{1}{c}{} & \multicolumn{3}{c}{HD 6582} &$|$& \multicolumn{3}{c}{HD 22879}&$|$& \multicolumn{3}{c}{HD 84937}&$|$& \multicolumn{3}{c}{HD 103095}\\
\hline
  El  &  1&  2& 3&$|$& 1&  2&  3&$|$& 1&  2&  3&$|$& 1&  2&  3  \\  	                            
\hline                                                                                                              
[Fe/H]&	--0.83&	--0.89	&0.06	&$|$&--0.91	&--0.85&--0.06&$|$&--2.24	&--2.08	&--0.16	&$|$&--1.35	&--1.34	&--0.01  \\ 
                                                                                       
[Mg/H]&	--0.42&	--0.45	&0.03	&$|$&--0.59	&--0.48&--0.11&$|$&--1.82	&--1.76&--0.06&$|$&--1.06	&--1.14&0.08   \\  
                                                                                       
[Si/H]&	--0.52&	--0.58	&0.06	&$|$&--0.61	&--0.59&--0.02&$|$&--1.60	&--1.73&0.13	&$|$&--1.15	&--1.15&0.00 \\  
                                                                                       
[Ca/H]&	--0.59&	--0.57	&--0.02 &$|$&--0.65	&--0.53&--0.12&$|$&--1.78	&--1.67&--0.12&$|$&--1.07	&--1.24&0.17  \\  
                                                                                       
[Ti/H]&	--0.48&	--0.52	&0.04	&$|$&--0.61	&--0.55&--0.07&$|$&--1.77	&--1.66&--0.11&$|$&--0.97	&--1.24&0.27   \\ 
                                                                                       
[Sc/H]&	--0.54&	--0.69	&0.15	&$|$&--0.86	&--0.79&--0.07&$|$&--2.19	&--1.90&--0.30&$|$&--1.26	&--1.26&0.00   \\ 
                                                                                       
[V/H] &	--0.67&	--0.66	&--0.01 &$|$&--0.77	&--0.73&--0.04&$|$&--	&--	&--	&$|$&--1.20	&--1.40&0.20   \\ 
                                                                                       
[Cr/H]&	--0.85&	--0.83	&--0.03 &$|$&--0.95	&--0.86&--0.09&$|$&--2.23	&-2.23	&--0.00&$|$&--1.34	&--1.55&0.21   \\ 
                                                                                       
[Mn/H]&	--0.93&	--1.01	&0.08	&$|$&--1.16	&--1.16&--0.00&$|$&--	&--	&--	&$|$&--1.43	&--1.79&0.36   \\
                                                                                       
[Co/H]&	--0.65&	--0.72	&0.07	&$|$&--0.69	&--0.74&0.05  &$|$&--2.04	&--	&--	&$|$&--1.27	&--1.38&0.11   \\ 
                                                                                       
[Ni/H]&	--0.83&	--0.83	&--0.00 &$|$&--0.84	&--0.85&0.01  &$|$&--2.15	&--2.06&--0.09&$|$&--1.36	&--1.50&0.14   \\      
\hline
\end{tabular}                                         
\end{table*}

The mean values of $<$$\Delta$[El/Fe]$>$  are equal to  0.04 $\pm0.05$(HD6582), $-$0.05 $\pm0.05$(HD22879), $-$0.09 $\pm0.12$ (HD84937), 0.14 $\pm0.13$ (HD103095).  The greatest shift and spread are due to the difference in temperature obtained in these two studies.                                                       

The mean differences of abundance values (for this paper and the work of other authors) and rms deviations $<\Delta[El/Fe]>$ are in Table \ref{abun_liter}.

\begin{table}
\caption{The comparison of our abundance determinations with those of other authors: $<\Delta[El/Fe]>$ is mean differences of abundance values 
and rms deviations. }
\label{abun_liter}
\begin{tabular}{lcl}
\hline          
  HD     &  $<\Delta[El/Fe]>$      &   references      \\
\hline
 6582    &     -0.01$\pm0.10$  &    \cite{fulb:00}     \\
         &      0.04 $\pm0.07$  &      \cite{gratton:03}   \\
6833     &      -0.15$\pm0.20$  &    \cite{fulb:00}     \\ 				
19445    &      -0.04$\pm0.08$  &    \cite{fulb:00}     \\
         &      0.12$\pm0.22$   &     \cite{roed:14}        \\
         &      0.07$\pm0.17$ &       \cite{gratton:03}    \\
22879    &      -0.02$\pm0.06$  &    \cite{fulb:00}     \\
         &      0.07$\pm0.09$  &      \cite{klo:11}      \\
         &      0.06$\pm0.12$  &      \cite{gratton:03}     \\
84937    &      0.01$\pm0.10$  &    \cite{fulb:00}     \\
         &      0.05$\pm0.11$  &      \cite{gratton:03}    \\
103095   &      0.01$\pm0.11$  &    \cite{fulb:00}     \\
         &      0.09$\pm0.10$  &       \cite{gratton:03}   \\ 
216143   &		 -0.03$\pm0.16$  &    \cite{fulb:00}     \\
221170   &      0.04$\pm0.13$  &    \cite{fulb:00}     \\
         &     0.00$\pm0.13$   &     \cite{ivans:06}     \\
224930   &      0.00$\pm0.14$  &    \cite{fulb:00}     \\
         &      0.06$\pm0.09$  &      \cite{ston:12}     \\
         &      0.04$\pm0.08$ &        \cite{gratton:03}   \\ 
\hline
\end{tabular}                                         
\end{table} 

As can be seen from Table \ref{abun_liter}, there is good agreement between our results and those of other authors. The largest discrepancies correspond to differences between our stellar parameters with respect to those from the studies, by \cite{fulb:00} 
for HD6833 and \cite{roed:14} for HD19445. It should be noted that other stars' characterised values from \cite{fulb:00} corroborate with those in the present study within the stated error definitions. 
The star HD6833 is a star with CN-weak molecular lines which has no scaled solar chemical composition, but is characterised by the Na and Al deficit relative to the Mg and O abundances. If we compare the abundances of these elements obtained in this study and in \citep{fulb:00, luck:91} (Table \ref{abun_namgal}), we can see that the ratio of these elements is similar, while the metallicity value is different.
The comparison of the equivalent widths of the lines measured by us and by \cite{fulb:00}, $<$(EW(fulb) $-$ EW(our)$>$ = $-$ 1.83 $\pm5.65$, showed a good agreement between the values. The fact that \cite{fulb:00} have only used three lines of neutral iron for the HD6833 study while from 30 to 60 lines were used for other stars in their study, is the most plausible cause for this difference. The shift and scatter of values for HD19445 is due to the difference between the Al abundance obtained by us ([Al/Fe] = 0.19) and that one by \cite{roed:14} ([Al/Fe] = $-$0.56). If the comparison is made without accounting the Al abundance, we obtained: $<\Delta[El/Fe]>$ = 0.08$\pm0.13$. The difference in the Al abundance is due to the fact that \cite{roed:14} analysis did not take into account NLTE corrections, that at this metallicity is about 0.6 dex for the lines used.

\begin{table}
\caption{For HD6833, the comparison of our O, Na, Mg, Al abundance determinations with 
those of other authors: 1 -  \citet{fulb:00}, 2 - \citet{luck:91}.}
\label{abun_namgal}
\begin{tabular}{lccc}
\hline 
El     &  this work &  1   &  2   \\ 
\hline
[Fe/H] & --0.77  & --1.04   &  --0.75   \\
$[O/Fe]$ &  0.08  &  -   &    --0.21  \\
$[Na/Fe]$ &--0.37  &  --0.06&   --0.43   \\
$[Mg/Fe]$ & 0.04  &  0.45 &    0.15   \\
$[Al/Fe]$ & --0.30 & 0.16  &   --0.36  \\  
\hline
\end{tabular}                                         
\end{table}

\section{Results and discussion}
\label{sec: result, discuss}
 
Our results for different elements are summarised in Figure\ref{abfig} and in Table \ref{abund}. 
Among our target  stars there are two stars with peculiar chemical composition, HD6833, a CN-weak giant  and HD221170,  a rich $r$-process  metal-poor star. 

\textbf{HD6833}. For this star we have a chemical composition that is slightly different from solar scaled: there are under abundances for Na, Al, Mn, and Cu; also a small deficit of Sr and Y; and a significant excess of Ca and Eu compared to the Sun. 
The values of Ca, Mn and Eu correspond to those of these elements at this metallicity, Na and Al abundances depart significantly from the general trend.
As shown above, the ratio of these element abundances agrees with that obtained by \cite{luck:91}. At the same time, \cite{luck:91} showed that the CNO abundance in the CN - weak giants differs only slightly (within the definition error) from that of ``normal'' giants, and of giants with G-weak band, and also on standard calculations of stellar evolution. However, they stressed that there is still the problem of a small C deficit. 
The distinctive ratio of O and Na, Mg and Al abundances may serve as a test for theories of stellar evolution \citep[see e.g][]{deniss:96, deniss:98, prantzos:07, deniss:15},  including the stars with moderate deficit of Fe. Thus, this star with particular enrichment in some elements requires a special study. 
Therefore, in this study applied to GCE, we exclude this star.  

\textbf{HD221170}. It is a well known halo star with $r$-process enrichment. Also in our study, we took a star HD216143 with the similar parameters as for star HD221170,  to compare the chemical composition of these two stars. For HD221170 we have obtained a slight excess of Eu ($r$-process element) abundance relative to those for HD216143, and also of other elements formed in neutron capture processes. Overabundance of $r$-process elements in HD221170 is due to anomalous enrichment of pristine material from where the star formed, possibly indicating an incomplete mixing at that time \citep[e.g.][]{ivans:06}. We have not included HD221170 in our study applied to GCE.

\subsection{Membership of stars to galactic populations }

The necessary and sufficient criterion to classify each star into the thin disc, thick disc and halo of the Galaxy does not exist. However, with galactic velocities or orbital elements, metallicity and relative abundance of some chemical elements, it can be attempted to classify each star into its most probable stellar population.
For instance, \cite{hawkins:15} explore the Galactic disk-halo transition region between -1.20 $<$ [Fe/H] $<$ -0.55 and show that may be able to chemically label the Galactic components in a clean and efficient way independent of kinematics using [$\alpha$/Fe], [C+N/Fe], [Al/Fe], and [Mg/Mn]. 
Also, using the total spatial velocity or eccentricity of the orbit, it is possible to distinguish stars from the initial and accreted halo \citep{venn:04, carollo:10}.

\begin{figure}
\begin{tabular}{c}
\includegraphics[width=8cm]{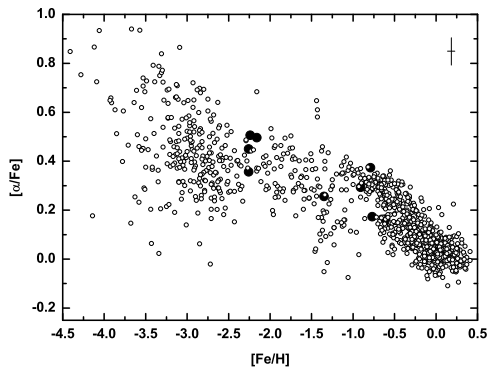}\\
\end{tabular}
\caption{Dependence of $\alpha$ elements abundance on [Fe/H].The elements Mg, Si and Ca were taken to calculate the averaged $\alpha$-element abundances. Our data marked as full circles and data of \citep{bensby:14, cohen:13, ishigaki:12, roederer:14} marked as small open circles.}
\label{alfa}
\end{figure}

Here we attempt to classify the target stars into the halo and thick disc populations according to their dynamics and abundances.
According to their high eccentricity (ecc $>$ 0.8, see Table \ref{kinpar}), HD6833, HD84937, HD103095, HD216143 are likely halo stars. 
HD84937 and HD216143 have also [Fe/H] $<$$-$2 which confirms their halo membership. HD6833 has a higher metallicity, [Fe/H]=$-$0.77, with a moderate $\alpha$ enhancement, 
[$\alpha$/Fe]=+0.17. 
In this work, the average abundance of Mg, Si and Ca are taken to calculate 
the [$\alpha$/Fe] ratio. The resulting $\alpha$ value is consistent with 
\cite{nissen:10}, where the 'low-$\alpha$' stars are suggested to be accreted from dwarf galaxies. However, HD6833 is a CN-weak star with peculiar chemical composition. 
According to \cite{hawkins:15}, this is a signature of the accreted halo.
HD224930 and HD170153 have more circular orbits confined close to the galactic plane which makes them more likely thick disc stars. 
HD224930 and HD170153 have V velocities, respectively -76 and +45 km/s, are not typical of thin disc. In the solar neighbourhood the thin disc rotates at about -10 km/s with respect to the Sun with a typical standard deviation of $\sim$20 km/s \citep[see for instance][]{soubiran:03}.
In addition both stars have [Fe/H] $<$ -0.50 which is also a characteristics of the thick disc. However their membership to the thin disc cannot be totally ruled out.
HD224930 has [Fe/H]=$-$0.79 and [$\alpha$/Fe] = 0.37 which is also typical of the thick disc. There are four stars (HD6582, HD19445, HD22879, HD221170) which have orbital parameters compatible either with the halo or the thick disc.  HD19445 and HD221170 have a low metallicity, [Fe/H] $<$$-$2, and a high $\alpha$ enhancement ([$\alpha$/Fe] $>$ +0.4) typical of the halo. HD6582 and HD22879 are intermediate in their kinematical and chemical properties which makes them impossible to classify. 
The dependence of [$\alpha$/Fe] with respect to [Fe/H] are shown in Figure \ref{alfa}. Membership in the galactic populations is given in Table \ref{kinpar}.

\subsection{A special element: Lithium}
  
Li is easily destroyed at typical H-burning conditions in stars. On the other hand, it may be produced as a result of extra-mixing processes in Asymptotic Giant Branch (AGB) stars and Red Giant Branch stars of different initial masses via the Cameron-Fowler transport mechanism \citep[see e.g.,][]{lattanzio:99,sackmann:99,nollett:03,denissenkov:11,palmerini:11}. Such
non-standard mixing processes are challenging to simulate for baseline one-dimensional stellar models, as large differences do exist in theoretical predictions \citep[e.g.,][]{lattanzio:15}.
Also because of this high dependence on the local stellar conditions, Li is a powerful diagnostic for stellar evolution, 
GCE models and cosmology. Li is made in the Big Bang \citep{bbfh:57}. Encouraging results for metal-poor stars which confirm Big Bang nucleosynthesis simulations were reported in the study by \cite{spite:82}. Metal-poor dwarfs studied by \cite{spite:82} showed similar Li abundances with small dispersion. This was referred to as the cosmological Li contribution. 
A large number of following investigations focused on both the cosmological Li abundance and the dispersion of that value, on both observational and theoretical ground \citep[see][and references therein]{cyburt:16}. In particular, a spread of the Li abundances was later found by \cite{thorn:93} and \cite{melendez:10}. The WMAP mission also confirmed that the cosmological Li abundance differs significantly from observations in metal-poor dwarfs. 
Among our stars, HD19445 and HD84937 with [Fe/H] $<$ $-$2.0 dex have long history of study $^{6}$Li/$^{7}$Li \cite[e.g.][]{smith:93, hobs:94, cayrel:99}. 
The isotopic ratio of $^{6}$Li/$^{7}$Li was proposed as an important indicator of efficiencies of mixing processes in the stellar interior. The Li formation in the solar spots \citep{livsh:97} and spallation reactions \cite{goriely:08}, and in the case of metal-poor stars, also for resolving of cosmological Li problem \cite[e.g.][]{asplund:06, fields:11}. However, recently \cite{lind:13} found that the observational support for significant $^{6}$Li production in the early Universe proposed by \cite{asplund:06} is substantially weakened by their findings.

\begin{figure}
\begin{tabular}{c}
\includegraphics[width=8cm]{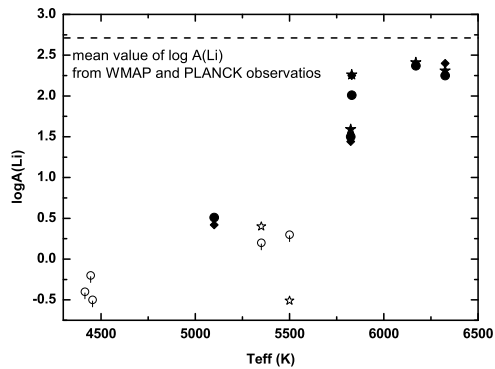}\\
\end{tabular}
\caption{Dependences of logA(Li) on \Teff. Our Li abundance and the upper limit  are marked as full and open circles, respectively.
Li abundance and  the upper limit are marked as full and open asterisks \citep[][]{ramirez:12}, and observations by \citep{fulb:00} with diamonds. The mean values of log A(Li) based on WMAP and PLANCK observations: 2.65 \citep{steigman:07},  2.72 \citep{cyburt:08} 
and 2.66 \citep{cyburt:16}.}
\label{li_t}
\end{figure}

In this work we provide the Li elemental abundance, but we do not provide the $^{6}$Li/$^{7}$Li ratio. For Li the isotope shift is small, and in order to confidently determine $^{6}$Li/$^{7}$Li we would need a spectrum with a resolution of about 100,000 and a ratio of signal to noise of 500.
In general, we obtain low Li abundance for stars with \Teff\ $<$ 5600 K. This supports the hypothesis of the destruction of Li by H-burning depleting the pristine Li concentration (see Table \ref{li} and Figure \ref{li_t}). 
From our stellar sample, the star HD 19445 shows a Li abundance = 2.1 that is 0.08 dex lower than the average Li abundance. This is the largest variation that we observe, that is within observational errors. 
For all the other stars we get variations lower than 0.04 dex.

At present, the GCE of Li in the Galaxy is uncertain, due to the fact that Li is not only easily destroyed in stellar interiors during the stellar evolution, but can also be produced by stars as mentioned above. To avoid any possible Li abundance variations caused by stellar evolution, only dwarf stars should be used 
\citep[with \Teff\ $>$ 5600 K and \logg $>$ 3.7, e.g.,][]{guiglion:16}. We have four stars with such parameters, namely, HD 19445, HD22879, HD84937, and HD170153. Among them there are two stars with [Fe/H] $<$ -1.5, HD19445 and HD22879, that have the values of [Fe/H] and logA(Li) close to these values for the Spite plateau (logA(Li) = 2.2) found by \cite{spite:82}.  
Taking into account the Standard Big Bang Nucleosynthesis model, the primordial Li abundance is predicted to be logA(Li) = 2.6 \citep[][]{spergel:03}. Li can be produced in the interstellar medium via spallation by Galactic cosmic rays (GCR) and by different types of stars \citep[see e.g.,][]{guiglion:16}. 
The chemical evolution of Li in the Milky Way was constructed by \cite{prantzos:12}. According to these results, GCR and primordial nucleosynthesis can produce at most $\sim$30 percent of solar Li, but its stellar production is too low to explain the missing Li component.

\subsection{Comparing observations with GCE simulations}
\label{sec: element nucleosynthesis}

The evolution of the chemical inventory of the galaxy from its early stages \citep[e.g.,][]{sneden:08,bonifacio:12,griffen:16} until the youngest stellar generations observed in open clusters and associations \citep[e.g., ][]{biazzo:12,carraro:15,mishenina:15} provides an invaluable source of information about the galaxy formation, its evolution and about how real stars work. 
A consistent interpretation of the evolution of elemental ratios at different metallicites is one of the main task of GCE. GCE models are folding theoretical stellar yields within the fundamental physics equations driving the dynamics of the galaxy. This allows to test theoretical models, its chemical products and all the different pieces of physics relevant for a given observable. 
For instance, the observation of C and N in old stars (C and N abundances are not provided in this analysis) in the early galaxy may provide insights about the core-collapse supernova (CCSN) engine, how fast the massive star progenitors were rotating and about ingestion of material between different stellar layers, or about the C production in AGB stars, that are the most relevant source of C and N in the galactic disk and in the Sun \citep[e.g.,][]{spite:05,chiappini:06,bonifacio:15,pignatari:15,frischknecht:16,Yoon:16}. 
The lighter $\alpha$-elements O and Mg are indicative of the evolution of massive star progenitors \citep[e.g.,][]{thielemann:96}, while heavier $\alpha$ elements (Si, Ca, Ti) and the Fe group elements are mostly affected by the CCSN explosion at low metallicities \citep[e.g.,][]{woosley:95,woosley:02,thielemann:11b,  nomoto:13} and by SNe Ia once these start to contribute to the galactic chemical inventory \citep[e.g.,][and references therein]{hillebrandt:13}.
The chemical evolution of heavy elements allows to constrain theoretical simulations for the $s$-process \citep[e.g.,][]{kaeppeler:11}, the $r$-process \citep[][and references therein]{thielemann:11}, and for a wide range of less constrained nucleosynthesis processes, like the $i$-process \citep[e.g.,][]{cowan:77,herwig:11,bertolli:13,dardelet:15,mishenina:15,jones:15,roederer:16,hampel:16} and a zoo of explosive neutrino-wind components from Supernovae \citep[e.g.,][]{frohlich:06a,frohlich:06b,qian:08,farouqi:09,roberts:10,arcones:11,hansen:11,wanajo:11,arcones:13,hansen:13}.
However, the results from GCE simulations depend also on the assumptions and simplifications made by the model \citep[e.g.,][]{gibson:03}, and on the theoretical stellar yields adopted. 

In Figure \ref{abfig}, we compared the results obtained from our stellar sample, with observations from other stars in the Milky Way.
Within the observational errors, in general our stars show abundance patterns consistent with the average chemical enrichment history of the Milky Way.
In the same figure, we also provide a sample of prediction from GCE models, calculated using different codes, assumptions and stellar yields.
Departure of single stars from the average evolutionary trends of elemental ratios may be due to observational errors or peculiar enrichment histories. The stars in our sample follow quite well the average chemical evolution of the Milky Way. In the following part of the section, we therefore compare predictions from different GCE models with the average abundance trends in the galaxy. The conclusions that we will derive also apply to our stellar sample.

Black lines have been produced by OMEGA, a one-zone model that is available online with the NuGrid NuPyCEE chemical evolution package\footnote{https://github.com/NuGrid/NUPYCEE}. This simple code is designed to capture the global trends generated by a set of stellar yields and to provide an easy-to-use platform to test and compare stellar models.  It takes into account inflows of primordial gas and galactic outflows driven by star formation (see \citealt{cote16a}). The star formation history and the dark matter and gas masses are input parameters in order to mimic the evolution of a specific galaxy, here the Milky Way. OMEGA assumes homogenous mixing but takes into account the delay between star formation and stellar ejecta. Each stellar population formed throughout a simulation, using SYGMA (C. Ritter et al., in preparation), is followed in time by considering their specific age, mass, and metallicity. We refer to \cite{cote16b} for more details on the different input parameters for stellar populations and to \cite{cote16c} for the numerical setup for the Milky Way.

For OMEGA, we used NuGrid AGB stellar yields for low and intermediate-mass stars (\citealt{pignatari:16}; C. Ritter et al., in preparation).  Type Ia supernovae are included with a delay-time distribution in the form of $t^{-1}$ (\citealt{mmn14}) and the yields calculated by \cite{tny86}. The black solid and dashed lines in Figure \ref{abfig} have respectively been generated with the massive star yields of West \& Heger (private communication) using the \cite{ertl16} and the \textit{no-cutoff} prescriptions for the stellar remnant masses (see \citealt{cote16c} for more details). The black dotted lines represent NuGrid Set 1 extension massive star yields (C. Ritter et al., in preparation), using the stellar remnant mass prescription of \cite{fbw12}, along with the zero-metallicity yields of West \& Heger (private communication).
Massive star yields are only applied to stars with initial mass between 8 and 30~M$_\odot$. We thus assume no ejecta for stars more massive than 30~M$_\odot$ (see discussion in \citealt{cote16b}). 
For the ejection of $r$-process material, we only considered the contribution of neutron star mergers.  We used the delay-time distribution of the standard models of \cite{dominik12} to distribute the yields as a function of time for each simple stellar population. We assumed that each neutron star merger ejects a total mass of 0.01\,M$_\odot$ with the $r$-process composition provided by \cite{arnould07}. Overall, our implementation generates $5.5\times10^{-5}$ neutron star merger event per unit of solar mass formed.

The GCE model predictions by \cite{bisterzo:14} are shown in Figure \ref{abfig} with red lines. 
This code was presented by \cite{travaglio:99,travaglio:04},  
and follows the composition of stars, stellar remnants, interstellar matter (atomic and molecular gas), and their mutual interaction, in the three main zones of the Galaxy, halo, thick disk, and thin disk. The chemical evolution is calculated inside the solar annulus, located 8.5 kpc from the Galactic centre. The thin disk is divided into independent concentric annuli. The chemical evolution is regulated by the star formation rate (SFR), initial mass function (IMF), and nucleosynthesis yields from different stellar  mass ranges and populations. The star formation rate has been determined
self-consistently as the result of aggregation, interacting and interchanging processes of the interstellar gas, which may occur spontaneously or stimulated by the presence of other stars. The treatment of the elemental matrix and yields have been updated by \cite{bisterzo:14}, as presented here.

Concerning the heavy elements, the $r$-process yields are derived as explained by \cite{travaglio:99}.
Because of the large uncertainties affecting the astrophysical site and physical conditions of the r-process, the solar r-process contribution for elements heavier than Ba is derived by adopting the r-process residuals method \citep[e.g.,][]{arlandini:99}. This method is evaluated by subtracting the s-process contributions from the solar abundances (N$_{\rm r}$ = N$_\odot$ -- N$_{\rm s}$), still providing a good approximation to derive the solar r-process abundances from Ba to Bi.
We assumed the r-process yields as primary and occurring in core-collapse supernovae with a limited range of progenitor masses (8-10 M$_\odot$ supernovae), following the observed decreasing trend of heavy neutron-capture elements in the early Galaxy.

As discussed by \cite{travaglio:04}, we have included an additional primary contribution (LEPP) to interpret the trend observed for light neutron-capture elements (as Sr-Y-Zr).

Concerning massive stars, for this work we took the yields from: 1) massive stars from 13 to 30 M$_{\odot}$ from West, Heger et al. (private communication) with the no-cutoff remnant mass prescription; 2) stars more massive than 30M$_{\odot}$ and up to 100 M$_{\odot}$ from \cite{limongi:12} and \cite{chieffi:13} 
(up to Mo); 3) Type Ia supernovae from \cite{travaglio:05}. 

The green solid line corresponds to the model predictions associated with the solar neighbourhood chemical evolution model described by \cite{hughes:08}, realised with the GEtool software package \citep[][]{fenner:03}. Nucleosynthetic yields were drawn from \cite{woosley:95}, \cite{karakas:07}, and \cite{nomoto:97}  
for CCSNe, AGB stars, and Type Ia supernovae, respectively. Linear extrapolation of the CCSNe yields from 40 M$_\odot$ to 60 M$_\odot$ was employed, with a lower mass limit of 0.08 M$_\odot$ adopted, with the mass and metallicity interpolation scheme as outlined by \cite{gibson:97}; 
the distribution of stellar masses employed in the modelling was that described by \cite{kroupa:93}. 
A dual infall framework was used with a rapid initial infall of gas on essentially a free-fell timescale (50 Myrs), referred to as the halo phase, followed by a more protracted infall phase on an exponential timescale of 10.5 Gyr (after a delay of 500 Myr with respect to the halo phase). A conservative star formation law predicated upon the class Schmidt Law was employed with the star formation  rate proportional to the square of the local gas surface density, modulated by an efficiency factor of 0.06 Gyr$^{-1}$. 
The overall model is constrained to recover a local total mass surface density of 55 M$_\odot$/pc$^2$ in the solar neighbourhood.
The $r$-process yields were simply estimated using the residual method, i.e., from the difference between solar and $s$-process predictions \citep[e.g.,][]{arlandini:99}. 

Finally, the results for the inhomogeneous chemical evolution model "ICE" are shown with magenta cross symbols for O, Mg, Si, Ca and Eu.
ICE is able to keep track of the intrinsic inhomogeneities in the interstellar medium. In comparison to other GCE models which employ an instantaneous mixing approximation, the inhomogeneities in our model produce a scatter in observed abundances, especially at lower metallicities, before a sufficient number of events cause convergence to average values.
The main iteration procedure of one time step ($10^6$ years) can be summarised as follows.
Primordial gas is falling into the simulation volume. The star formation rate is calculated via a Schmidt-Kennicutt law with a power of $1.5$. Cells are chosen randomly to trigger star formation, however, higher density cells are favoured. The mass of a newly born star is chosen randomly from a Salpeter initial mass function (with an integrated slope of $-1.35$). The newly born stars inherit the abundances of the local interstellar medium. The life time of a star is calculated via an age-life expectance relation of the Geneva Stellar Evolution and Nucleosynthesis Group (e.g., \citealt{Schaller92}). 
Dependent on their initial mass, stars which reached the end of their life time will either undergo a CCSN event (with stellar yields given by \citealt{Thielemann96}) or blow off processed matter into the interstellar medium via a planetary nebula. A fraction of $9 \cdot 10^{-4}$ of all intermediate star binary systems will results in a type Ia supernova explosion, with an ejecta composition taken from \cite{Iwamoto99} (model CDD2 yields). A fraction of $3.8 \cdot 10^{-4}$ of high mass star binary systems ends in a neutron star merger event after both stars have independently exploded as CCSNe and an inspiral delay time (or coalescence time) of 10 million years has passed. The ejecta of these events are taken from \cite{Korobkin12}, following a solar $r$-process distribution. Stars in the surrounding interstellar medium are polluted by the ejecta of the respective nucleosynthesis event.

The main difference to the other models presented here including $r$-process element yields (e.g., for Eu), is that 
ICE assumes that in addition a fraction of all CCSNe (0.1 per cent) explodes as "magneto-rotationally driven supernovae", leading to the formation of magnetars, i.e. neutron stars with magnetic fields beyond 10$^{15}$ Gauss (see \citealt{Winteler12}, or \citealt{Nishimura15} for discussion), and producing $r$-process elements in polar jet ejecta during the explosion. 
Thus, the main difference is that also an $r$-process source exists related to massive single stars which does not require the delay of binary evolution after individual SN explosions producing Fe and the merger event producing $r$-process elements.

A detailed description of the chemical evolution model can be found in \cite{Wehmeyer15}.
While GEtool, OMEGA and three-zone models (green, black and red lines, respectively) represent spatially averaged values of the abundance scatter
observed at low metallicities, the approach adopted by ICE model (pink crosses) provides a more realistic view of the local chemical
inhomogeneities detected in the interstellar medium at early times.
ICE predictions allow to study not only the average trend for a given elemental ratio with metallicity, but also the dispersion at any given time due to local inhomogeneities, before the stellar products are fully mixed. These inhomogeneities allow to explain the scatter in elemental abundances, especially at low metallicities. With the largest scatter seen in $r$-process elements, the biggest advantage of an inhomogeneous GCE treatment is revealed: When a rare sub-class of supernovae ("magneto-rotationally driven supernovae", with an occurrence rate of probably less than 1\% with respect to regular CCSNe) pollutes its environment, the $r$-process elemental ratio is extremely high in comparison to regions where such a pollution did not take place. This inhomogeneity effect especially at low(est) metallicities might thus be an explanation for the observed large scatter in $r$-process elemental abundances.
A key element to test the inhomogeneous halo issue is Eu. In particular, the ICE model
suggests that different nucleosynthesis sources (neutron star mergers and fast rotating CCSNe driven by high magnetic fields) are needed to
reproduce the large Eu spread observed in the Galactic halo \citep[][]{Wehmeyer15}. 

We have seen in Section \ref{sec: stellar param}, that spectroscopic observations for the same star may differ from one analysis to the other for several reasons. 
In the same way, different results can be obtained also for theoretical GCE predictions.
While for some elements (e.g., Mn) the theoretical GCE results are quite close to each other, for most of the elements large variations are obtained.
In particular, the OMEGA prediction using the NuGrid yields (black dotted lines) show the largest departures from other models, and for many cases (e.g., Ni and Zn) from the observations. 
The larger [Co/Fe], [Ni/Fe], [Cu/Fe], [Zn/Fe] and [Zr/Fe] are due to the contribution from the $\alpha$-rich freezout component \citep[e.g.,][]{woosley:92,pignatari:16}, present in CCSNe models from stellar progenitors with initial mass 12 and 15 M$_{\odot}$. The [Cr/Fe] bump at solar-like metallicities is only due to the yields of the 20 M$_{\odot}$ star model, that are affected by the merger of the C and O shells. 

Taking into account the large variation between the different predictions, all the models cannot reproduce the observed trend for [Sc/Fe], [Ti/Fe] and [V/Fe].
These issues are well known, and there is not yet a clear solution at least for TI and V 
\citep[e.g.,][]{kobayashi:11,sneden:16}. 
Concerning Sc, it was shown by \citep[][]{frohlich:06a} that when neutrino interactions with the innermost ejected zones are treated correctly, Sc underabundance in the CCSN ejecta is strongly reduced. This is due to the effect that neutrinos increase the electron fraction Ye slightly above 0.5. However, none of the presently existing yield tables for GCE studies take into account of these results. This issue could indicate that in real CCSNe a range of entropies and electron fractions are obtained the ejected matter, which can be only obtained within multi-dimensional simulations.
Therefore, for these cases the presently available predictions of CCSN nucleosynthesis suffer the shortcomings that
none of them are based on self-consistent multi-dimensional explosion models. This leads to three types of problems:
(a) One-dimensional piston as well as thermal bomb methods utilise assumed explosion energies of the order 1-1.2 $10^{51}$ erg. This does not reflect differences in the pre-explosion stellar
models, e.g. changing compactness, and therefore a variation in the expected range of explosion energies and mass cuts, related strongly to $^{56}$Ni and other Fe-group ejecta. Hopefully more 
sophisticated upcoming approaches like PHOTB or PUSH can solve this issue \citep{perego:15,sukhbold:16}.
(b) The presently utilised models do not include the effect of neutrino interactions with matter
deep in explosive layers. Neutrino and Anti-neutrino absorption on protons and neutrons leads to slightly proton-rich conditions, based on the neutron/proton mass difference for similar neutrino and antineutrino spectra and luminosities.
This can improve the predictions for Sc, Cu and Zn \citep{frohlich:06a,frohlich:06b}. 
(c) In addition, limitations of one-dimensional CCSN models, neglecting the role of convection (which is the key
aspect in realistic three-dimensional explosions) are affecting the robustness of nucleosynthesis results in particular for intermediate-mass and Fe group elements. 
Most likely, these limitations are already relevant in the final stages of stellar progenitor models, and are also related to the difficulties in obtaining robust explosions from the last generation of CCSN multi-dimensional simulations \citep[e.g.,][and references therein]{muller:16}.

For Cr, Mn, and Ni, the predictions from the three-zone code (red lines) are similar to the ones of OMEGA (black solid and dashed lines). This better agreement compared to light $\alpha$ elements such as Mg implies that Cr, Mn, and Ni are not significantly affected by the ejecta of stars more massive than 30 M$_\odot$, as OMEGA did not include them.
The large scatter seen for elements heavier than Zn at low [Fe/H] between the GCE models is caused by the different assumptions used for the $r$-process and the $s$-process. 
Concerning the $r$-process elements (e.g., Eu), in the three-zone (red lines), GEtool (green lines), and ICE (pink crosses) codes, some CCSNe contribute to the evolution and provide a short-timescale enrichment that allows an early appearance of predictions on the [Fe/H] axis. On the other hand, the OMEGA code (black lines) only considered the contribution of neutron star mergers, which require a certain delay before contributing to the chemical evolution of heavy elements. The ICE model also considered the contribution of neutron star mergers, but an additional earlier $r$-process source is included in the simulations \citep[][]{Wehmeyer15}.  
The variations seen between the black lines are only caused by different Fe yields associated with different choices of massive star models, as the same number of neutron star mergers and the same $r$-process yields were used. The choice of stellar yields can thus have an impact on the interpretation of how many $r$-process events is needed in order to explain the observations, at least when Fe is the element of reference in the abundance ratios. 
Concerning the prediction for typical $s$-process elements (e.g., Ba), the predictions by OMEGA are lower than observations. This is due to the adopted $s$-process yields from AGB models. For these simulations, the convective-boundary mixing mechanism assumed to be responsible for the formation of the $^{13}$C pocket are producing a weaker $s$-process production compared to the most $s$-process rich AGB stars observed in the galactic disk, and compared to measurements in pre-solar mainstream SiC grains \citep[][]{pignatari:16}. These GCE results confirm these conclusions over the galactic metallicity range, and provide additional insights about physics processes relevant for stellar physics.

The ICE model (pink crosses), at least for O, Mg, Si, Ca, and Eu, suggests that non-uniform mixing in the interstellar medium can generate scatter in the predictions that is larger than the scatter seen in the different components of the three-zone code, which means that data can be reproduced without implying the Galactic structure. On the other hand, the three-zone code can address the formation history of the different components of our Galaxy, which cannot be done with the ICE model.
Secondly, using different stellar yields can also lead to large differences in the theoretical GCE simulations. 
For instance, it is still controversial the role of hypernovae to explain observations of iron-group elements \citep[e.g.,][]{nomoto:13,sneden:16}, while the impact of using different stellar yields is so relevant \citep[see e.g.,][and the discussion in this section]{romano:10,molla:15}. The controversial GCE role of different types of SNe~Ia can also be considered as a similar source of uncertainty at [Fe/H] above $\sim-1$ \citep[e.g.,][]{seitenzahl:13,mishenina:15a}.
However, at present none of these scenarios have been established and a definitive solution for Sc, Ti and V still needs to be found.
For heavy elements, there are even more uncertainties to consider. The existence of a large variety of nucleosynthesis mechanisms highlighted from observations in the early galaxy \citep[e.g.,][]{roederer:10,hansen:12,roederer:16,frischknecht:16} in principle does not provide strong constraints about their effective relevance for the chemical inventory of the Sun \citep[][]{travaglio:04,bisterzo:14,trippella:14,cristallo:15}. Nevertheless, there are strong observational indications now that the nucleosynthesis paradigm where the abundances beyond Fe are just made by a sum of $s$-process and $r$-process need to be revised \citep[e.g.,][]{mishenina:15}.

\section{Conclusions}
\label{sec: conclusions}

In this work, we presented and discussed the abundance measurements of 10 stars, with metallicity $-2.2 <$ [Fe/H] $< -0.6$. 
The same objects have been previously analysed by other authors, using different spectroscopic lines, methods, and assumptions. 
For most studied stars the observed abundances for all the elements are consistent between all the authors. 
The largest discrepancy is obtained for star HD19445 for Al, whose abundance was determined with and without consideration of NLTE corrections. 
Some discrepancy for HD103095 between our determinations and \cite{jofre:14} is due to the difference in temperature obtained in these two studies. 
And also, the  discrepancy for the peculiar star HD6833 is possibly due to the small number of iron lines used to 
determine the metallicity by \cite{fulb:00}.
In our case the O abundance has the largest error, ranging between 0.10 and 0.2 dex, it is due to the O weak lines that we used. The
best measured abundances are of Cr, Fe, Mn, the errors are between 0.03 and 0.11 dex.

We have compared the observations with an extended sample of predictions from GCE simulations. 
The study of the origin of the elements is based on the comparison between observations and GCE predictions. 
However, while stellar observations have usually provided with a clear error analysis, 
uncertainty in the theoretical GCE results need also to be considered. The main sources of this uncertainty are from stellar yields and 
from different assumptions in GCE simulations, among others, the stellar mass range on which stellar yields
are applied, the interpolation scheme between stellar models, the stellar initial mass function, the star formation history, the star
formation efficiency (related to the gas fraction), the treatment of SNe~Ia, the astrophysical sites for heavy elements, and the
galaxy framework (single- or multi-zone). 
Different GCE models produce a scatter larger than observational errors for many elements. Furthermore,
all these theoretical simulations are not consistent with the observed chemical evolution of the elemental ratios [Sc/Fe], [Ti/Fe] and [V/Fe]. While for Ti and V a clear solution has not been found yet, a promising scenario to solve the Sc problem has been discussed by \cite{frohlich:06a}, but the impact of neutrino-winds ejecta still need to be tested within a GCE context.
These problems not new, and here we can confirm them by using the results of four GCE codes. 
In particular, we considered six GCE models, including different theoretical stellar yields and a large variety of assumptions.
We highlight that the present theoretical stellar yields from CCSNe are most likely the dominant source of this discrepancy between theory and observations, which is one of the most important puzzle that modern multi-dimensional CCSN simulations will need to solve. 
This underlines that improved CCSN nucleosynthesis predictions from realistic models are required. Among others, this requires a detailed study of the progenitor stellar structure in the last days before the SN explosion, the role of rotation and magnetic fields, the effect of neutrino-interactions on the innermost ejected layers, and more substantially the role of the multi-dimensional explosion character. 
This and the possible role of hypernovae events for more massive stars, are prerequisites to a more comprehensive representation of the chemical evolution of the Galaxy.

\section*{Acknowledgements}

We thank the anonymous referee for very helpful comments and suggestions.
TM, TG, MP, FKT and SK thank for the support from the Swiss National 
Science Foundation, project SCOPES No. IZ73Z0$_{}$152485. MP 
acknowledges significant support to NuGrid from NSF grants PHY 09-22648 
(Joint Institute for Nuclear Astrophysics, JINA), NSF grant PHY-1430152 
(JINA Center for the Evolution of the Elements) and EU 
MIRG-CT-2006-046520. MP acknowledges the support from SNF (Switzerland). 
BC acknowledges financial support from the FRQNT (Quebec, Canada) 
postdoctoral fellowship program. FKT acknowledges support from the Swiss 
SNF and the European Research Council (FP7) under ERC Advanced Grant 
Agreement 321263 FISH. SB thanks JINA (ND Fund 202476) for financial 
support. BKG acknowledges the support of STFC, through grants 
ST/J001341/1 and ST/G003025/1. Numerical calculations have been 
supported by access to facilities, including the B2FH Association 
(http://www.b2fh.org/) and the University of Hull's High Performance 
Computing Facility, \it viper\rm.

\bibliographystyle{mnras}
\bibliography{metpoorstars}

\appendix
\section{}

\begin{table}
\caption{Atomic data for the lines used in synthetic method calculations. NLTE calculations marked  in Table.}
\label{Synth}
\begin{tabular}{lcccccc}
\hline
El     & $\lambda$& log gf  & Elow  & Note\\ 
       &  0.1nm   &         & eV    &     \\
\hline                                 
O I     & 6300.304 & -9.717 & 0.000 & NLTE \\ 
Na I    & 5682.630 & -0.708 & 2.102 & NLTE \\ 
Na I    & 5688.190 & -1.407 & 2.104 & NLTE \\ 
Na I    & 5688.200 & -0.452 & 2.104 & NLTE \\ 
Na I    & 5889.940 & 0.108  & 0.000 & NLTE \\ 
Na I    & 5895.920 & -0.195 & 0.000 & NLTE \\ 
Na I    & 6154.220 & -1.560 & 2.102 & NLTE \\ 
Na I    & 6160.740 & -1.260 & 2.104 & NLTE \\ 
Mg I    & 4167.271 & -0.745 & 4.346 & NLTE \\ 
Mg I    & 4702.990 & -0.440 & 4.346 & NLTE \\ 
Mg I    & 4730.020 & -2.292 & 4.346 & NLTE \\ 
Mg I    & 5172.680 & -0.451 & 2.712 & NLTE \\ 
Mg I    & 5183.600 & -0.240 & 2.717 & NLTE \\ 
Mg I    & 5528.400 & -0.498 & 4.346 & NLTE \\ 
Mg I    & 5711.080 & -1.720 & 4.346 & NLTE \\ 
Mg I    & 6318.710 & -1.839 & 5.108 & NLTE \\ 
Mg I    & 6319.230 & -2.060 & 5.108 & NLTE \\ 
Mg I    & 6319.490 & -2.537 & 5.108 & NLTE \\ 
Al I    & 3944.006 & -0.622 & 0.000 & NLTE \\ 
Al I    & 3961.520 & -0.322 & 0.014 & NLTE \\ 
Al I    & 5557.063 & -2.377 & 3.143 & NLTE \\ 
Al I    & 6696.023 & -1.479 & 3.143 & NLTE \\ 
Al I    & 6698.673 & -1.780 & 3.143 & NLTE \\ 
Ca I    & 4108.526 & -0.824 & 2.709 & NLTE \\ 
Ca I    & 4226.728 & 0.244  & 0.000 & NLTE \\ 
Ca I   &  4283.011 & -0.220 & 1.886 & NLTE \\ 
Ca I   &  4289.367 & -0.300 & 1.879 & NLTE \\ 
Ca I   &  4302.528 & 0.280  & 1.899 & NLTE \\ 
Ca I   &  4318.652 & -0.211 & 1.899 & NLTE \\ 
Ca I   &  4355.079 & -0.420 & 2.709 & NLTE \\ 
Ca I   &  4425.437 & -0.360 & 1.879 & NLTE \\ 
Ca I   &  4434.957 & -0.010 & 1.886 & NLTE \\ 
Ca I   &  4435.679 & -0.523 & 1.886 & NLTE \\ 
Ca I   &  4454.779 & 0.260  & 1.899 & NLTE \\ 
Ca I   &  4512.268 & -1.892 & 2.526 & NLTE \\ 
Ca I   &  4526.928 & -0.548 & 2.709 & NLTE \\ 
Ca I   &  4578.551 & -0.697 & 2.521 & NLTE \\ 
Ca I   &  4685.268 & -0.880 & 2.933 & NLTE \\ 
Ca I   &  5188.844 & -0.115 & 2.933 & NLTE \\ 
Ca I   &  5260.387 & -1.719 & 2.521 & NLTE \\ 
Ca I   &  5261.704 & -0.579 & 2.521 & NLTE \\ 
Ca I   &  5265.556 & -0.114 & 2.523 & NLTE \\ 
Ca I   &  5349.465 & -0.310 & 2.709 & NLTE \\ 
Ca I   &  5512.980 & -0.464 & 2.933 & NLTE \\ 
Ca I   &  5581.965 & -0.555 & 2.523 & NLTE \\ 
Ca I   &  5588.749 & 0.358  & 2.526 & NLTE \\ 
Ca I   &  5590.114 & -0.571 & 2.521 & NLTE \\ 
Ca I   &  5594.462 & 0.097  & 2.523 & NLTE \\ 
Ca I   &  5601.277 & -0.523 & 2.526 & NLTE \\ 
Ca I   &  5857.451 & 0.240  & 2.933 & NLTE \\ 
Ca I   &  5867.562 & -1.570 & 2.933 & NLTE \\ 
Ca I   &  6102.723 & -0.770 & 1.879 & NLTE \\ 
Ca I   &  6122.217 & -0.319 & 1.886 & NLTE \\ 
Ca I   &  6162.173 & -0.090 & 1.899 & NLTE \\ 
Ca I   &  6163.755 & -1.286 & 2.521 & NLTE \\ 
Ca I   &  6166.439 & -1.143 & 2.521 & NLTE \\ 
Ca I   &  6169.042 & -0.797 & 2.523 & NLTE \\ 
Ca I   &  6169.563 & -0.478 & 2.526 & NLTE \\ 
Ca I   &  6439.075 & 0.390  & 2.526 & NLTE \\ 
\hline                                                 
\end{tabular} 
\end{table}                                                                            

\begin{table}
\contcaption{}
\label{Synth}
\begin{tabular}{lcccccc}
\hline
El     & $\lambda$& log gf  & Elow  & Note\\ 
       &  0.1nm   &         & eV    &     \\
\hline                                 
Ca I   &  6462.567 & 0.262  & 2.523 & NLTE \\ 
Ca I   &  6471.662 & -0.686 & 2.526 & NLTE \\ 
Ca I   &  6493.781 & -0.109 & 2.521 & NLTE \\ 
Ca I   &  6499.650 & -0.818 & 2.523 & NLTE \\ 
Ca I   &  6572.779 & -4.296 & 0.000 & NLTE \\ 
Ca I   &  6717.681 & -0.523 & 2.709 & NLTE \\ 
Sc II  &  4670.400 & -0.580 & 1.357 &      \\ 
Sc II  &  5239.823 & -0.770 & 1.455 &      \\ 
Sc II  &  5318.336 & -2.040 & 1.357 &      \\ 
Sc II  &  5526.770 & 0.130  & 1.768 &      \\ 
Sc II  &  5657.886 & -0.500 & 1.507 &      \\ 
Sc II  &  5667.000 & -1.240 & 1.500 &      \\ 
Sc II  &  5669.038 & -1.120 & 1.500 &      \\ 
Sc II  &  5684.190 & -1.080 & 1.507 &      \\ 
Sc II  &  6245.621 & -0.980 & 1.507 &      \\ 
Sc II  &  6604.582 & -1.480 & 1.357 &      \\ 
Mn I   &  4502.221 & -0.345 & 2.920 &  HFS \\ 
Mn I   &  4709.720 & -0.340 & 2.890 &  HFS \\ 
Mn I   &  4739.113 & -0.490 & 2.941 &  HFS \\ 
Mn I   &  4754.039 & -0.086 & 2.282 &  HFS \\ 
Mn I   &  4761.530 & -0.138 & 2.953 &  HFS \\ 
Mn I   &  4762.375 & 0.425  & 2.889 &  HFS \\ 
Mn I   &  4783.420 & 0.042  & 2.300 &  HFS \\ 
Mn I   &  4823.514 & 0.144  & 2.320 &  HFS \\ 
Mn I   &  5432.550 & -3.795 & 0.000 &  HFS \\ 
Mn I   &  6013.497 & -0.251 & 3.073 &  HFS \\ 
Mn I   &  6021.803 & 0.034  & 3.075 &  HFS \\ 
CuI    &  5105.545 & 1.390  & -1.510&  HFS \\ 
CuI    &  5218.209 & 3.820  & 0.270 &  HFS \\ 
CuI    &  5782.136 & 1.640  & -1.780&  HFS \\ 
Sr II  &  4077.709 & 0.167  & 0.000 & NLTE \\ 
Sr II  &  4161.792 & -0.501 & 2.940 & NLTE \\ 
Sr II  &  4215.519 & -0.144 & 0.000 & NLTE \\ 
Y II   &  4374.933 & 0.271  & 0.409 &      \\ 
Y II   &  4398.010 & -0.894 & 0.130 &      \\ 
Y II   &  4854.876 & -0.110 & 0.990 &      \\ 
Y II   &  4883.682 & 0.265  & 1.080 &      \\ 
Y II   &  4900.119 & 0.103  & 1.033 &      \\ 
Y II   &  4982.129 & -1.289 & 1.033 &      \\ 
Y II   &  5087.416 & -0.169 & 1.084 &      \\ 
Y II   &  5119.112 & -1.359 & 0.992 &      \\ 
Y II   &  5200.406 & -0.569 & 0.992 &      \\ 
Y II   &  5205.722 & -0.192 & 1.033 &      \\ 
Y II   &  5402.774 & -0.629 & 1.839 &      \\ 
Ba II  &  4554.034 & 0.163  & 0.000 &HFS, NLTE\\ 
Ba II  &  5853.675 & -1.000 & 0.604 &HFS, NLTE\\ 
Ba II  &  6141.714 & -0.076 & 0.704 &HFS, NLTE\\ 
Ba II  &  6496.900 & -0.377 & 0.604 &HFS, NLTE\\ 
La II  &  4086.710 & -0.069 & 0.000 &      \\ 
La II  &  4123.236 & 0.110  & 0.321 &      \\ 
La II  &  4238.391 & -0.219 & 0.403 &      \\ 
La II  &  4526.097 & -0.649 & 0.772 &      \\ 
La II  &  4558.460 & -0.969 & 0.321 &      \\ 
La II  &  4662.509 & -1.239 & 0.000 &      \\ 
La II  &  4716.440 & -1.209 & 0.772 &      \\ 
La II  &  4748.720 & -0.539 & 0.927 &      \\ 
La II  &  4921.790 & -0.449 & 0.244 &      \\ 
La II  &  4986.830 & -1.299 & 0.173 &      \\ 
La II  &  5122.989 & -0.849 & 0.321 &      \\ 
La II  &  5163.612 & -1.809 & 0.244 &      \\ 
La II  &  5290.840 & -1.649 & 0.000 &      \\ 
La II  &  5303.530 & -1.349 & 0.321 &      \\ 
La II  &  5482.270 & -2.229 & 0.000 &      \\ 
\hline                                                 
\end{tabular} 
\end{table}                                                                            

\begin{table}
\contcaption{}
\label{Synth}
\begin{tabular}{lcccccc}
\hline
El     & $\lambda$& log gf  & Elow  & Note\\ 
       &  0.1nm   &         & eV    &     \\
\hline                                 
La II  &  5808.310 & -2.199 & 0.000 &      \\ 
La II  &  6390.480 & -1.409 & 0.321 &      \\ 
La II  &  6774.000 & -1.819 & 0.126 &      \\ 
PrII   &  4222.950 & 0.235  & 0.055 & HFS  \\ 
PrII   &  4408.820 & 0.053  & 0.000 & HFS  \\ 
PrII   &  4510.150 & -0.007 & 0.422 & HFS  \\ 
PrII   &  5259.730 & 0.114  & 0.633 & HFS  \\ 
PrII   &  5322.770 & -0.123 & 0.482 & HFS  \\ 
Nd II  &  4018.810 & -0.849 & 0.064 &      \\ 
Nd II  &  4021.340 & -0.099 & 0.321 &      \\ 
Nd II  &  4069.260 & -0.390 & 0.064 &      \\ 
Nd II  &  4368.630 & -0.809 & 0.064 &      \\ 
Nd II  &  4446.380 & -0.349 & 0.205 &      \\ 
Nd II  &  4462.980 & 0.040  & 0.559 &      \\ 
Nd II  &  4501.810 & -0.689 & 0.205 &      \\ 
Nd II  &  4706.540 & -0.709 & 0.000 &      \\ 
Nd II  &  4797.150 & -0.689 & 0.559 &      \\ 
Nd II  &  4825.480 & -0.419 & 0.182 &      \\ 
Nd II  &  4947.020 & -1.129 & 0.559 &      \\ 
Nd II  &  4959.120 & -0.799 & 0.064 &      \\ 
Nd II  &  5092.790 & -0.609 & 0.380 &      \\ 
Nd II  &  5249.580 & 0.200  & 0.976 &      \\ 
Nd II  &  5255.510 & -0.669 & 0.205 &      \\ 
Nd II  &  5293.160 & 0.100  & 0.823 &      \\ 
Nd II  &  5306.460 & -0.969 & 0.859 &      \\ 
Nd II  &  5311.450 & -0.419 & 0.986 &      \\ 
Nd II  &  5319.810 & -0.139 & 0.550 &      \\ 
Nd II  &  5356.970 & -0.279 & 1.264 &      \\ 
Nd II  &  5485.700 & -0.119 & 1.264 &      \\ 
Nd II  &  5533.820 & -1.229 & 0.559 &      \\ 
Nd II  &  5548.450 & -1.269 & 0.550 &      \\ 
Sm II  &  4188.125 & -0.440 & 0.543 &      \\ 
Sm II  &  4424.321 & 0.140  & 0.484 &      \\ 
Sm II  &  4434.320 & -0.070 & 0.378 &      \\ 
Sm II  &  4452.722 & -0.410 & 0.277 &      \\ 
Sm II  &  4467.341 & 0.150  & 0.659 &      \\ 
Sm II  &  4499.475 & -0.870 & 0.248 &      \\ 
Sm II  &  4511.830 & -0.820 & 0.184 &      \\ 
Sm II  &  4523.909 & -0.390 & 0.439 &      \\ 
Sm II  &  4536.512 & -1.280 & 0.104 &      \\ 
Sm II  &  4577.688 & -0.650 & 0.248 &      \\ 
Sm II  &  4642.228 & -0.460 & 0.378 &      \\ 
Eu II  &  4129.720 & 0.220  & 0.000 &      \\ 
Eu II  &  6645.060 & 0.120  & 1.380 &      \\ 
Gd II  &  4085.558 & -0.010 & 0.731 &  \\ 
Gd II  &  4130.366 & 0.139  & 0.731 &      \\ 
Gd II  &  4191.075 & -0.479 & 0.427 &  \\ 
Gd II  &  4215.022 & -0.439 & 0.427 &  \\ 
Gd II  &  4419.290 & -0.699 & 0.492 &  \\ 
\hline                                                 
\end{tabular} 
\end{table}                                                                            

\onecolumn

\clearpage
\begin{longtable}{llccc}                                                     
\caption{Comparison of atmospheric parameters with the data of other authors.}
\label{compar}\\
\hline                                                                  
HD&\Teff &\logg	&[Fe/H]& sources  \\           
\hline                                                                  
\endfirsthead
\hline                                                                  
HD&\Teff &\logg	&[Fe/H]& sources  \\           
\hline                                                                  
\endhead
\hline                                                                  
   6582&  5308 & 4.41& --0.89&   \cite{jofre:14}   \\
		   &	5412&	4.56&	--0.8 &		\cite{ramirez:13}        \\	
       &	5526&	4.49&	--0.77&		\cite{gray:03}            \\      
       &	5240&	4.3&	--0.94&		\cite{mish:11}          \\      
       &	5291&	4.57&	--0.89&		\cite{maldon:12}          \\      
       &	5331&	4.54&	--0.81&		\cite{takeda:07}           \\      
       &	5387&	4.51&	--0.83&		\cite{zhao:00}          \\      
       &	5390&	4.45&	--0.83&		\cite{mash:00}          \\      
       &	5250&	4.40&	--0.98&		\cite{fulb:00}     \\      
       &	5240&	4.20&	--0.89&		\cite{mishkov:01}       \\ 
			&  5322& 4.46&	--0.82 	& \cite{gratton:03}  \\
      mean   &	5336 $\pm87$&	4.44 $\pm11$&	--0.86 $\pm0.06$&           \\     
 \hline  
 6833&		4450&	1.4&	--1.04&		\cite{fulb:00}        \\    
&	4400&	1.5&	--0.85&		\cite{mash:12}         \\    
&	4450&	1.4&	--1.04&		\cite{molenda:13}         \\    
&	4400&	1&	--0.89&		\cite{mishkov:01}       \\    
mean&	4425 $\pm29$&	1.32 $\pm$0.22&	--0.96 $\pm0.10$&	                       \\    
\hline			                                     
 19445&	5820&	3.65&	--2.28&	\cite{roed:14}          \\
 &	6055&	4.43&	--1.83&		\cite{ramirez:13}       \\
&	5920&	4.3&	--1.98&		\cite{gray:03}           \\
&	5982&	4.38&	--2.13&	\cite{hansen:13}          \\
&	5890&	4.48&	--2.12&	\cite{sozz:09}          \\
&	6016&	4.43&	--1.95&	\cite{	zhao:00}         \\
&	6020&	4.38&	--1.95&	\cite{mash:00}          \\
&	6047&	4.51&	--1.96&	\cite{gratton:00}         \\
&	5825&	4.20&	--2.13&	\cite{fulb:00}     \\
&	5890&	4.48&	--2.12&	\cite{molenda:13}          \\
&	6136&	4.43&	 7.35*& \cite{vanden:14}    \\
&	6000&	4.00&	--1.89&	\cite{mishkov:01}         \\
& 6135	&4.46&	--2.01& \cite{cas:10}  \\
&5890&	4.50&	--2.04&  \cite{klo:11}  \\
&5976&	4.44&	--2.04&  \cite{gratton:03}  \\
mean&	5973 $\pm99$&	4.34 $\pm0.23$&	--2.03 $\pm0.12$&	without  \cite{vanden:14}   \\
\hline                                                    
 22879& 5786& 4.23& --0.85 & \cite{jofre:14} \\
&	5970&	4.52&	--0.81& \cite{bensby:14}           \\ 
  &	5949&	4.68&	--0.79&	 \cite{tsantaki:13}        \\ 
&	5910&	4.30&	--0.83&	\cite{ramirez:13}        \\ 
&	5884&	4.52&	--0.82&	\cite{adibekyan:12}	  \\ 
&	5759&	4.25&	--0.85&	\cite{nissen:11}          \\ 
&	5972&	4.5&	--0.77&	\cite{	mish:11}          \\ 
&	5827&	4.45&	--0.69&	\cite{sozz:09}          \\ 
&	5774&	4.20&	--0.86&	\cite{nissen:00}         \\ 
&	5870&	4.27&	--0.86&	\cite{	mash:00}          \\ 
&	5800&	4.30&	--0.91&	\cite{	fulb:00}     \\ 
&	5857&	4.46&	--0.83&	\cite{sousa:08}         \\ 
&	5775&	4.26&	--0.83&	\cite{yan:15}          \\ 
&	5800&	4.29&	--0.84&	\cite{sitn:15}                   \\ 
&	5941&	4.41&	--0.91&	\cite{cas:10}           \\ 
&5802	&4.37&	--0.78&  \cite{klo:11}  \\
&5827&	4.44&	--0.79&  \cite{gratton:03}  \\
mean &	5853 $\pm73$&	4.37 $\pm0.13$&	--0.83 $\pm0.05$&			        \\ 
\hline                                                     
84937& 6275  & 4.11 & --2.08&   \cite{jofre:14} \\
 & 	6541&	4.23&	--1.92&	\cite{bensby:14}        \\
     &	6431&	4.08&	--2.15&		\cite{ishigaki:12}         \\
&	6377&	4.15&	--2.02&		\cite{ramirez:13}       \\
&	6206&	3.89&	--2.20&		    \cite{boesg:11}       \\
&	6350&	4.03&	--2.07&	   \cite{mash:00}        \\
&	6375&	4.1&	--2.08&	     \cite{fulb:00}   \\
&	6431&	4.08&	--2.15&	\cite{lind:13}       \\
&	6350&	4.09&	--2.12&	    \cite{sitn:15}               \\
&	6408&	3.93&	--2.11&	   \cite{cas:10}          \\
&	6300&	4.00&	--2.15&	\cite{lawler:13}         \\
&	6250&	3.80&	--2.00&		\cite{mishkov:01}     \\
& 6290&	4.02&	--2.18&  \cite{gratton:03}  \\
mean &	6353 $\pm90$&	4.04 $\pm0.11$&	--2.09 $\pm0.08$&			      \\
 \hline                                                 
 103095& 4827&	4.6&	--1.34& \cite{jofre:14} \\
& 5149&	4.71&	--1.27&		\cite{ramirez:13}      \\ 
      &	5157&	4.76&	--1.08&		\cite{gray:03}         \\ 
&	5144&	4.05&	--1.12&		\cite{maldon:12}     \\ 
&	5095&	4.79&	--1.29&		\cite{takeda:07}        \\ 
&	5014&	4.75&	--1.44&		\cite{sozz:09}        \\ 
&	5110&	4.67&	--1.35&		\cite{	zhao:00}        \\ 
&	5110&	4.66&	--1.35&		\cite{mash:00}        \\ 
&	5152&	4.77&	--1.17&		\cite{gratton:00}     \\ 
&	4950&	4.50&	--1.46&		\cite{	fulb:00}    \\ 
&	5000&	4.40&	--1.39&		\cite{mishkov:01}   \\
&	5130&	4.66&	--1.26&	         \cite{sitn:15}          \\
&5025&	4.63&	--1.28&  \cite{gratton:03}  \\
    mean&	5066 $\pm99$&	4.61 $\pm0.20$&	--1.29 $\pm0.12$&                        \\ 
\hline
 170153&6173&	4.22&	--0.58&		\cite{ramirez:13}    \\       
 &	6034&	4.28&	--0.65&		 \cite{chen:00}        \\       
mean&	6104 $\pm98$&	4.25 $\pm0.04$&	--0.62 $\pm0.05$&                       \\       
 \hline                                                       
 216143&       	4525&	1&	--2.25&		\cite{	fulb:00}     \\     
&	4525&	0.80&	--2.18&		\cite{burris:00}        \\     
&	4525&	1.00&	--2.25&		\cite{molenda:13}         \\     
&	4529&	1.30&	--2.1&		         \cite{ishigaki:14}             \\     
mean&	4526 $\pm2$&	1.03 $\pm0.21$&	--2.20 $\pm0.07$&	                       \\     
 \hline                                                       
 221170 	&4500&	0.9&	--2.19&		\cite{fulb:00}  \\	      
	&4425&	1.00&	--2.15&		\cite{burris:00}       \\     
	&4510&	1.00&	--2.16&		\cite{mash:12}       \\     
	&4444&	0.92&	--2.12&		\cite{molenda:13}      \\     
	&4510&	1.00&	--2.09&		\cite{ivans:06}       \\     
	&4475&	1.00&	--2.09&	 	\cite{yush:05}	       \\     
	&4500&	1.00&	--2.05&		\cite{mishkov:01}  \\      
mean	&4481 $\pm34$&	0.97 $\pm0.04$&	--2.12 $\pm0.05$&	                       \\     
\hline                                                        
 224930&	5510&	4.46&	--0.76&		\cite{ramirez:13}    \\ 
&	5502&	4.27&	--0.64&		\cite{gray:03}         \\ 
&	5300&	4.10&	--0.91&		\cite{mish:11}	           \\ 
&	5491&	4.75&	--0.72&		\cite{maldon:12}            \\ 
&	5680&	4.86&	--0.52&		\cite{takeda:07}         \\ 
&	5275&	4.10&	--1.00&		        \cite{	fulb:00}   \\ 
&	5357&	4.32&	--0.9&		        \cite{molenda:13}         \\ 
&	5480&	4.45&	--0.66&			\cite{yan:15}   	           \\ 
&	5300&	4.10&	--0.85&		\cite{mishkov:01}   	   \\
&5470	& 4.20&	--0.71 & \cite{ston:12}    \\
&5357&	4.32&	--0.87&  \cite{gratton:03}  \\
mean&	5429 $\pm122$&	4.36 $\pm0.26$&	--0.78 $\pm0.14$&	                           \\ 
\hline                                                        
\hline                                                                                           
\end{longtable}

\label{lastpage}

\bsp

\end{document}